%% file: main.tex
\documentclass[sigconf, nonacm, screen]{acmart}

\usepackage{fontawesome5}
\usepackage[dvipsnames]{xcolor}
\usepackage{soul}
\usepackage{amsmath}
\usepackage{multirow}
\usepackage{graphicx}
\usepackage{tikz}
\usetikzlibrary{arrows.meta,positioning,fit,calc,shapes.geometric,shapes.symbols}
\usepackage{array}
\usepackage{pdflscape}

\DeclareMathAlphabet{\mathbbold}{U}{bbold}{m}{n}

\definecolor{SpecGreen}{RGB}{0,170,20} 
\definecolor{SpecTeal}{RGB}{0,120,140} 
\definecolor{SpecBlue}{RGB}{0,50,160} 
\definecolor{SpecPurple}{RGB}{150,60,170}
\definecolor{SpecRed}{RGB}{200,0,0}

\AtBeginDocument{%
  }

\begin{document}

\title{A Systematic Multi-Domain Evaluation of Document Retrievers}

\author{Valentin Velev}
\orcid{0009-0006-5098-154X}
\affiliation{%
  \institution{University of Konstanz}
  \city{Konstanz}
  \country{Germany}
}
\email{valentin.velev@uni-konstanz.de}

\author{Andreas Spitz}
\orcid{0000-0002-5282-6133}
\affiliation{%
  \institution{University of Konstanz}
  \city{Konstanz}
  \country{Germany}
}
\email{andreas.spitz@uni-konstanz.de}

\renewcommand{\shortauthors}{Valentin Velev and Andreas Spitz}


\begin{abstract}
    Document retrieval is a crucial component of many modern AI systems, directly influencing their effectiveness, robustness, and fairness in downstream tasks. While recent years have seen a growing number of retrievers, comparative studies in the literature are typically limited in scope or focused on singular benchmarks, domains, or model families. This fragmentation makes it difficult to draw reliable conclusions about the relative strengths, weaknesses, and trade-offs of document retrievers.
    To address this gap, we conduct a large-scale empirical evaluation of document retrievers, covering three families (sparse, dense, and expansion-based) and evaluating 33 retrievers across seven IR datasets, analyzing retrieval quality, runtime, and failure points.
    Rather than tuning each model individually, we evaluate every retriever off the shelf, under the configuration reconstructable from its public documentation and a uniform compute budget.
    Our results show that NV-Embed-v2 achieves the strongest performance on four of the seven datasets, albeit at the cost of substantial query latencies. Among sparse retrievers, we find that SPLADE-v3 rivals the top-performing approach despite much lower latency, and even achieves top scores on MS MARCO. On instruction-following datasets, \textsc{GritLM} delivers the best performance.
    Finally, an analysis of the retrievers' failure points reveals contrasts between models and families that indicate potential for unrealized gains in retrieval performance.
    \begin{center}
    \faGithub\quad\url{https://github.com/valentinsvelev/retriever-evaluation}
    \end{center}
\end{abstract}



\maketitle

\section{Introduction}

Document retrieval, as a subset of information retrieval (IR), involves identifying and ranking textual documents according to a user's information need \citep{manning2008ir}. Nowadays, IR systems underpin a wide range of applications used in everyday life spanning web search (e.g., Google Search, Bing Search), recommender systems (e.g., Netflix, Amazon, Spotify), as well as more recent contributions to natural language processing, such as retrieval-augmented generation (RAG) and agentic AI systems utilizing large language models (LLMs). Consequently, document retrieval is no longer a standalone application but a critical dependency, whose behavior directly affects the performance, reliability, and fairness of the systems that depend on it. The past decade has seen significant progress in retrieval methodology, driven primarily by advances in neural architectures capable of learning contextualized dense, low-dimensional vector representations of text \citep{guo2022irsurvey, hambarde2023ir}.

Despite these advances, document retrieval remains a challenging task. Dense retrieval methods have been shown to be sensitive to linguistic variations in queries \citep{cao2025writing, chari2023spelling, chari2025transliteration}, specific document types and their content \citep{fayyaz2025collapse}, as well as biases introduced during training \citep{macavaney2022abnirml}. Furthermore, single-vector dense retrievers, i.e., the majority of modern retrievers, share the fundamental representational inability to fully represent all combinations of top-$k$ documents unless they utilize a sufficiently large embedding dimension \citep{weller2026limit}. In other words, single-vector retrievers struggle to determine which documents to retrieve due to the number of possible retrieval targets, with the only feasible solution being a scaling-up of the model size, thereby sacrificing query latency for retrieval performance.

While novel approaches to addressing these problems are proposed frequently, we unfortunately find that new retrievers tend to be evaluated on just a small fraction of the available benchmarks that do not represent the full breadth of application settings and document domains. Furthermore, query latencies are only seldom considered, despite the ever-growing sizes of neural models that should necessitate such an inquiry, in particular to enable a grounded comparison to classic or more lightweight sparse approaches.
Our study addresses this gap by systematically evaluating representative sparse, dense, and expansion-based models across a diverse set of retrieval tasks.

Specifically, our aim is to \textbf{measure the retrieval effectiveness obtainable under the configurations readily available to a practitioner} using a model's public artifacts, subject to a uniform compute budget. We explicitly avoid tuning each retriever to its optimum on individual datasets to \textbf{ensure comparability} between the tested retrievers on a level playing field, rather than providing best-case performances a user would not be able to realistically replicate in practice.

In summary, our \textbf{contributions} are fourfold:
\begin{itemize}
    \item[$(1)$] We empirically evaluate the retrieval performance of 33 high-profile document retrievers on seven benchmarks, including 63\% new results.
    \item[$(2)$] For the 37\% of results that have previously been reported in the literature, we include a detailed discussion of reproducibility and comparability hurdles.
    \item[$(3)$] We evaluate query latencies to provide an accurate perspective on the models' performance-runtime trade-offs.
    \item[$(4)$] Finally, we investigate the failure points of all retrievers by comparing the queries that they fail to rank consistently.
\end{itemize}

\section{Background and Related Work}

We provide an overview of the underlying paradigms of retrievers that we consider as well as an overview of related surveys and benchmarks, before discussing the selection of models we evaluate.

\subsection{Retrieval Paradigms}

\textbf{Sparse Retrieval.} 
Early retrieval systems, \textit{lexical retrievers}, inferred relevance from the terms shared between a query and a document, giving rise to sparse retrievers that compute relevance using corpus statistics (e.g., term frequency). 

Modern sparse retrieval extends this paradigm through \textit{learned sparse retrievers}. In classic models such as BM25 \citep{robertson2009bm25}, term relevance is computed from corpus statistics, whereas in learned sparse retrievers it is derived from predicted (context-aware) term weights \citep{nguyen2023lsr}. In both cases, queries and documents can be viewed as high-dimensional sparse vectors, where non-zero elements correspond to terms that contribute to the score, enabling efficient retrieval via inverted indexes such as Lucene.

\textbf{Dense Retrieval.}
Instead of relying on lexical overlap, dense retrievers encode queries and documents using language models trained with contrastive or distillation-based objectives, producing low-dimensional contextualized embeddings $\mathbf{h}_q$ for queries and $\mathbf{h}_d$ for documents \citep{zhao2024dense}. Relevance is then estimated as a $\mathrm{score}(q,d) = g(\mathbf{h}_q, \mathbf{h}_d)$, where $g(\cdot)$ is a similarity function, often cosine similarity. Unlike sparse retrieval, this score cannot be expressed as a sum of independent term-level contributions and thus requires approximate nearest neighbor (ANN) search for efficient retrieval, typically implemented using indexes like FAISS \citep{douze2024faiss}.

While dense retrievers differ in retrieval mechanism, backbone architecture, pooling strategies, and training paradigm, they can be grouped based on the underlying network architecture into \textit{encoder-only retrievers} that produce contextualized embeddings, \textit{encoder-decoder retrievers} that are generalizable and can be instruction-aware, and \textit{decoder-only retrievers} based on generative models, typically also with instruction-following capabilities yet repurposed for retrieval. In the following, we use this categorization of dense retrievers (see Table \ref{tab:retrievers}). Finally, dense retrievers also differ with regard to their training, with some models being trained specifically for retrieval using contrastive learning objectives, whereas others are trained on instruction-following or for creating general embeddings used beyond retrieval, such as clustering or classification.

\textbf{Expansion-based Retrieval.}
Unlike sparse and dense retrieval methods, which primarily rely on matching sparse lexical or dense semantic representations, expansion-based retrieval methods augment queries or documents with additional tokens, typically using a generative language model. The main objective of this approach is to mitigate the negative impact of distribution shifts \citep{weller2024survey}. We refer the reader to the survey by \citet{li2026queryexpansion} for a detailed discussion.

\subsection{Related Work}

\textbf{Surveys.}
Prior work has mapped the shift from classic sparse methods to neural dense retrievers and hybrid architectures \citep{guo2022irsurvey, hambarde2023ir}, surveyed dense retrieval techniques based on pre-trained language models \citep{zhao2024dense}, reviewed architectural developments up to modern LLM-based and generative retrieval systems \citep{li2025survey, xu2026survey, zhu2026llm4ir}, and surveyed query expansion methods \citep{li2026queryexpansion}. These surveys characterize model families and design trends but do not offer a systematic comparison of retrievers. In contrast, we empirically evaluate retrievers under consistent experimental conditions.

\textbf{Benchmarking Studies} 
evaluate selected subsets of retrieval models, and prior work has examined API-based embedding models and their effectiveness relative to baseline retrievers \citep{kamalloo2023api}, analyzed learned sparse retrievers across BEIR datasets \citep{thakur2023sprint}, and conducted an in-depth analysis of specific benchmarks such as BEIR's Touché-2020 \citep{thakur2024beir}. Others investigate query and document expansion methods \citep{weller2024survey}, out-of-distribution robustness \citep{liu2025ood}, multilingual retrieval \citep{oro2025comparison}, knowledge-intensive retrieval tasks \citep{abdallah2025comparing}, and deep research \citep{meng2026deep}. While these works offer insights within the respective model family or evaluation setting, they do not provide a systematic comparison of a range of retrievers on a range of datasets.

\begin{table*}[ht]
    \centering
    \caption{Evaluated retrievers grouped by family and architecture. Sparse retrievers are grouped by retrieval mechanism, dense retrievers by the type of language model they use.}

\vspace*{-5pt}
    
    \small
    \begin{tabular}{ll p{1.3\columnwidth}}
        \toprule
        \textbf{Family} & \textbf{Architecture} & \textbf{Retrievers} \\
        \midrule
        \multirow{2}{*}{Sparse} & Lexical & BM25 \citep{robertson2009bm25} \\
        & Learned & DeepCT \citep{dai2020deepct}, SPARTA \citep{zhao2021sparta}, uniCOIL \citep{lin2021unicoil}, SPLADE-v3 \citep{lassance2024spladev3} \\[2pt]
        \midrule
        \multirow{6}{*}{Dense} & \multirow{3}{*}{Encoder-only} & DPR \citep{karpukhin2020dpr}, ColBERTv2 \citep{santhanam2022colbertv2}, TAS-B \citep{hofstatter2021tasb}, ANCE \citep{xiong2021ance}, SimCSE \citep{gao2021simcse}, RetroMAE \citep{xiao2022retromae}, coCondenser \citep{gao2022cocondenser}, Contriever \citep{izacard2022contriever}, COCO-DR (large) \citep{yu2022cocodr}, \textsc{SimLM} \citep{wang2023simlm}, \textsc{Dragon}+ \citep{lin2023dragon}, mE5 (large) \citep{wang2024me5}, GTE (large v1.5) \citep{li2023gte}, BGE (large v1.5) \citep{xiao2024bge}, Jina Embeddings v3 \citep{sturua2025jina3}, Nomic Embed (v1.5) \citep{nussbaum2025nomic} \\
        & Encoder-decoder & GTR (large) \citep{ni2022gtr}, \textsc{InstructOR} (large) \citep{su2023instructor} \\
        & \multirow{2}{*}{Decoder-only} & LLM2Vec \citep{behnamghader2024llmvec}, RepLLaMA \citep{ma2024repllama}, \textsc{GritLM} \citep{muennighoff2025gritlm}, Qwen3 Embedding (0.6B) \citep{zhang2025qwen3embedding}, NV-Embed-v2 \citep{lee2025nvembed}, EmbeddingGemma \citep{vera2025embeddinggemma}, KaLM-Embedding (v2.5) \citep{zhao2026kalmembeddingv2} \\[2pt]
        \midrule
        Expansion-based & Query/Document expansion & docT5query \citep{nogueira2019doct5query}, query2doc \citep{wang2023query2doc}, HyDE \citep{gao2023hyde} \\
        \bottomrule
    \end{tabular}
    \label{tab:retrievers}
\end{table*}

\subsection{Retrievers}
\label{sec:retrievers}

We evaluate a diverse set of sparse, dense, and expansion-based retrievers representative of the current state-of-the-art. Our \textbf{inclusion criteria} are: (1) public availability of model weights, (2) widespread adoption within the IR community (over 100 citations as of submission date), and (3) competitive performance on established retrieval benchmarks, such as the MTEB Leaderboard on Hugging Face\footnote{\url{https://huggingface.co/spaces/mteb/leaderboard}}. In cases where a retriever has multiple variants, such as SPLADE, BGE, GTE, or Qwen3 Embedding, we selected only the best performing or the one offering a favorable balance between performance and computational efficiency. 
For expansion-based methods, our unit of analysis requires a further restriction. Most techniques in the query expansion literature~\cite{li2026queryexpansion} are augmentations of an underlying retriever rather than retrievers in their own right, and comprehensively assessing them would require crossing each technique with all retrievers. We therefore include three canonical methods, covering document expansion, query expansion, and hypothetical-document generation, as reference points for the family rather than as an exhaustive treatment of it.
Table~\ref{tab:retrievers} lists all models that we evaluate and Table \ref{tab-a:retrievers} includes detailed overviews of each retriever.

\subsubsection{Sparse Retrievers}
BM25~\cite{robertson2009bm25} scores documents from corpus statistics and serves as a baseline. Three learned sparse retrievers replace these statistics with predicted term weights: DeepCT~\cite{dai2020deepct} regresses contextualized term importance from BERT embeddings but performs retrieval with BM25; SPARTA~\cite{zhao2021sparta} learns token-level sparse representations scored over an inverted index; and uniCOIL~\cite{lin2021unicoil} scores lexical matches through the dot product of their contextualized embeddings. SPLADE-v3~\cite{lassance2024spladev3} additionally learns \textit{expansion-aware} weights over the full WordPiece vocabulary~\cite{devlin2019bert}.
 
\subsubsection{Dense Retrievers}
 
\paragraph{Encoder-Only.} The earliest dense retrievers are dual encoders trained contrastively, differing primarily in how negatives are obtained: DPR~\cite{karpukhin2020dpr} established the architecture, TAS-B~\cite{hofstatter2021tasb} adds balanced topic-aware query sampling with MarginMSE distillation, ANCE~\cite{xiong2021ance} draws global hard negatives from a dynamically refreshed ANN index, and \textsc{Dragon}+~\cite{lin2023dragon} applies diverse data augmentation to a RoBERTa~\cite{liu2019roberta} backbone. A second group targets the pre-training stage, adapting language models for retrieval before any supervised fine-tuning: SimCSE~\cite{gao2021simcse} through dropout-based augmentation, RetroMAE~\cite{xiao2022retromae} through masked auto-encoding, coCondenser~\cite{gao2022cocondenser} through corpus-aware structural readiness, \textsc{SimLM}~\cite{wang2023simlm} through a representation bottleneck, Contriever~\cite{izacard2022contriever} through contrastive learning with random cropping and MoCo, and COCO-DR~\cite{yu2022cocodr} through continuous contrastive (COCO) pre-training with implicit distributionally robust optimization (iDRO). A third group comprises general-purpose embedding models trained in multiple stages for use beyond retrieval, including clustering and classification: mE5 (large)~\cite{wang2024me5}, GTE (large v1.5)~\cite{li2023gte}, BGE (large v1.5)~\cite{xiao2024bge}, Jina Embeddings v3~\cite{sturua2025jina3}, and Nomic Embed (v1.5)~\cite{nussbaum2025nomic}. Departing from all of these, ColBERTv2~\cite{santhanam2022colbertv2} represents documents as multiple vectors and scores them by late interaction, using residual compression and $k$-means-based indexing.
 
\paragraph{Encoder-Decoder.} GTR (large)~\cite{ni2022gtr} repurposes the T5~\cite{raffel2020t5} encoder as a universal text encoder for both queries and documents, and \textsc{InstructOR} (large)~\cite{su2023instructor} extends it to encode a task instruction alongside the input text.
 
\paragraph{Decoder-Only.} Recent retrievers repurpose generative language models as text encoders. LLM2Vec~\cite{behnamghader2024llmvec} is an unsupervised approach for converting LLMs using bidirectional attention, masked next-token prediction, and a SimCSE objective, while RepLLaMA~\cite{ma2024repllama} fine-tunes LLaMA-2 7B~\cite{touvron2023llama2} and takes the hidden representation of an appended EOS token. \textsc{GritLM}~\cite{muennighoff2025gritlm} and NV-Embed-v2~\cite{lee2025nvembed} are both built on Mistral 7B~\cite{jiang2023mistral7b}: the former is instruction-tuned jointly for embedding and generation, the latter for embedding alone using a latent attention layer for pooling. A parallel line of work applies the same instruction tuning at far smaller scale, including Qwen3 Embedding (0.6B)~\cite{zhang2025qwen3embedding}, EmbeddingGemma~\cite{vera2025embeddinggemma} at 300M parameters, and KaLM-Embedding (v2.5)~\cite{zhao2026kalmembeddingv2} at 0.5B.
 
\subsubsection{Expansion-Based Retrievers}
docT5query~\cite{nogueira2019doct5query} expands documents by generating multiple queries for each and retrieving with BM25. query2doc~\cite{wang2023query2doc} instead expands the query, generating a pseudo-document and again retrieving with BM25, while HyDE~\cite{gao2023hyde} generates a hypothetical document and retrieves with the Contriever.

\section{Experimental Setup}

We evaluate the retrievers in a unified framework across the datasets described in Section~\ref{subsec:data} to ensure comparability across retrievers.

\subsection{Evaluation Metrics}
\label{subsec:metrics}

We evaluate all retrievers using standard metrics, following the conventions of each dataset. Results are reported \textit{at cutoff k} ($@k$). Note that for BEIR's ArguAna dataset, we exclude self-matches in line with common practice\footnote{\url{https://github.com/beir-cellar/beir/issues/14}}.

\subsubsection{nDCG and Robustness}
The normalized Discounted Cumulative Gain at cutoff $k$ (nDCG$@k$) \citep{jarvelin2002ndcg} evaluates retrieval quality by weighting relevance according to rank position and normalizing against the optimal ranking. 
For a query $q$,
\begin{equation}
    \mathrm{DCG}_q@k = \sum_{i=1}^{k} \frac{2^{\mathrm{rel}_i}-1}{\log_2(i+1)}, \quad\quad \mathrm{nDCG}_q@k = \frac{\mathrm{DCG}_q@k}{\mathrm{IDCG}_q@k}, 
\end{equation}
where $\mathrm{rel}_i$ is the relevance label of the document at rank $i$ and IDCG$_q@k$ is the maximum possible DCG for query $q$. 
A related measure is Robustness$@k$ \citep{oh2024instructir}, where identical queries with instruction variants are grouped together and the minimum nDCG$@k$ within each group is taken, capturing worst-case behavior under instruction variation.

\subsubsection{Precision and Recall} 
Another set of metrics often reported are Precision and Recall at cutoff $k$ (P@$k$, R@$k$). These capture the proportion of retrieved documents that are relevant (precision) and the proportion of relevant documents retrieved within the cutoff (recall). Formally, for a given query $q$: 
\begin{equation} 
\text{P}_q@k = \frac{1}{k} \sum^k_{i=1} \mathbbold{1}\{\text{rel}_i > 0\}, \quad\quad \text{R}_q@k = \frac{1}{R_q} \sum^k_{i=1} \mathbbold{1}\{\text{rel}_i > 0\},
\label{eq:precision-recall}
\end{equation}
where $R_q$ denotes the total number of relevant documents for query $q$ and $\mathbbold{1}\{\cdot\}$ is an indicator function, returning 1 if the document at rank $i$ is relevant and 0 otherwise.

\subsubsection{Success}
Success@$k$ \citep{santhanam2022colbertv2} indicates whether a query has at least one relevant document in the top-$k$, assigning a $1$ if it does and a $0$ otherwise.

\subsubsection{MRR}
Mean Reciprocal Rank at cutoff $k$ (MRR@$k$)~\cite{kantor2000mrr} computes the reciprocal of the rank of the first relevant document. It therefore rewards placing a relevant document as early as possible. Let $\mathrm{rank}_q$ denote the rank of the first relevant document for query $q$ from the set of all queries $\mathcal{Q}$:
\begin{equation}
\mathrm{MRR@k} = \frac{1}{|\mathcal{Q}|} \sum_{q \in \mathcal{Q}}
\begin{cases}
\frac{1}{\mathrm{rank}_q}, & \text{if } \mathrm{rank}_q \le k, \\
0, & \text{otherwise}.
\end{cases}
\end{equation}

\subsubsection{MAP} 
Mean Average Precision (MAP)~\cite{manning2008ir} averages, over all queries, the precision computed at each rank holding a relevant document, and thus reflects the quality of the ranking across all relevant documents rather than only the first. Let $\mathrm{P}_q@i$ denote precision at rank $i$ (i.e., $\mathrm{P}_q@k$ evaluated at $k=i$ in Equation \ref{eq:precision-recall}). The average precision (AP) for query $q$ and the mean average precision (MAP) across all queries $\mathcal Q$ is
\begin{equation}
    \mathrm{AP}_q = \frac{1}{R_q} \sum_{i=1}^{N} \mathrm{P}_q@i \cdot \mathbbold{1}\{\mathrm{rel}_i > 0\},
    \quad
    \mathrm{MAP} = \frac{1}{|\mathcal{Q}|} \sum_{q \in \mathcal{Q}} \mathrm{AP}_q.
\end{equation}

\subsubsection{$p$-MRR}
The pairwise MRR ($p$-MRR) \citep{weller2025followir} evaluates the effect of retrieval instruction changes. Let $RR_{\text{og}} = 1/R_{\text{og}}$ and $RR_{\text{new}} = 1/R_{\text{new}}$ denote reciprocal ranks under the original and modified instructions:
\begin{equation}
p\text{-MRR}(q)=
\begin{cases}
\frac{RR_{\text{og}}}{RR_{\text{new}}}-1, & R_{\text{og}}>R_{\text{new}}, \\[4pt]
1-\frac{RR_{\text{new}}}{RR_{\text{og}}}, & \text{otherwise}.
\end{cases}
\end{equation}

\subsection{Datasets}
\label{subsec:data}

To evaluate the retrievers, we rely on a comprehensive selection of widely used English datasets covering in-domain, out-of-domain, and instruction-following retrieval scenarios. For each category, we use two datasets with complementary characteristics to balance trade-offs such as scale versus judgment quality, binary versus graded relevance, broad domain coverage versus focused long-tail evaluation, and instruction diversity versus instruction specificity and provide a more robust assessment of retrieval performance. Each dataset consists of a set of queries, documents, and relevance judgments. For datasets in which documents include both a title and body, we prepend the title to the body. Similarly, for datasets with query instructions, they are appended to the query. The number of queries and documents for each dataset is available in Table \ref{tab-a:dataset-sizes}.

\subsubsection{MS MARCO and TREC-DL}
For in-domain evaluation and comparability to prior work, we use MS MARCO (dev set; without titles) \citep{bajaj2018msmarco}, TREC-DL 19 \citep{craswell2020trecdl2019}, and TREC-DL 20 \citep{craswell2021trecdl2020}. MS MARCO is a large-scale passage ranking dataset that includes web search queries paired with relevant passages from web documents. It is evaluated using MRR$@10$. The TREC-DL datasets use graded human-annotated qrels together with the MS MARCO documents and are commonly evaluated using nDCG$@10$.

\subsubsection{BEIR}
To assess out-of-domain (OOD) generalization, we rely on BEIR's \citep{thakur2021beir} 14 publicly available datasets, using Touché-2020 v2. BEIR is a widely-used multi-domain and multi-task benchmark. For evaluation, we use the standard nDCG$@10$.

\subsubsection{LoTTE}
In addition to BEIR, we use LoTTE \citep{santhanam2022colbertv2}, a benchmark derived from StackExchange posts. It emphasizes retrieval in long-tail scenarios, where queries may be rare, domain-specific, or outside the scope of general-purpose training data typically used for pre-training language models. To evaluate retrieval quality on LoTTE, we use Success$@5$.

\subsubsection{\textsc{InstructIR}}
We also include instruction-following benchmarks, as many modern retrievers are instruction-tuned. The first such dataset is \textsc{InstructIR} \citep{oh2024instructir}. It is derived from MS MARCO and introduces queries with multiple different user-aligned instructions and instruction-specific qrels. Following the authors, we report both Robustness$@10$ and nDCG$@10$.

\subsubsection{\textsc{FollowIR}}
As a second instruction-following benchmark, we consider \textsc{FollowIR} \citep{weller2025followir}, which includes three TREC datasets (News'21, Core'17, and Robust'04) paired with long, realistic instructions derived from TREC narratives, testing whether models appropriately adjust rankings when instruction constraints change. Its main metrics are MAP$@1000$, nDCG$@5$, and $p$-MRR.

\subsection{Implementation Details}
\label{sec:implementation}
\subsubsection{General}
To accommodate model-specific package dependencies, we use four separate Python environments as setup for our experiments. \textbf{Our code is available on GitHub}\footnote{
\url{https://github.com/valentinsvelev/retriever-evaluation}
}.

We determine the batch sizes per model based on throughput. All datasets are loaded using the \texttt{ir\_datasets} \citep{macavaney2021irdatasets} and \texttt{datasets} \citep{lhoest2021datasets} packages. Metrics are computed with \texttt{ir\_measures} \citep{macavaney2022irmeasures}.

For inference of all models, we use four GPUs with the exception of ColBERTv2, which we limited to a single GPU due to multi-GPU stability issues. Where supported, we employed available model-specific parallelization strategies for the remaining models, including data parallelism, model sharding across devices, and multiprocessing. For space and index efficiency, the embeddings and indexes for the largest models (LLM2Vec, RepLLaMA, \textsc{GritLM}, NV-Embed-v2) were created iteratively in batches of $2^{17}$.

\subsubsection{Sparse Retrievers}
\label{sec:sparse-ret}
We implement sparse models using the \texttt{Pyserini} package \citep{lin2021pyserini}. For BM25, we use Lucene’s flat index with default parameters ($k_1=0.9$, $b=0.4$), while DeepCT is configured with $k_1=10$ and $b=0.9$. Note that the authors of BEIR originally use a multi-field index \citep{thakur2021beir}, which produces different results\footnote{Our results match those for the flat Lucene index available at \url{https://castorini.github.io/pyserini/2cr/beir.html}}.
Learned sparse retrievers (i.e., SPLADE-v3, uniCOIL, and DeepCT) are instantiated from their official Hugging Face checkpoints, and for each document we extract the models' predicted token-level weights. These scores are then rescaled (e.g., multiplied by $100$), rounded to integers, and pruned by retaining only the highest-weighted terms (top-$k$ filtering with $k=512$) \citep{qiao2023thresholding, lassance2023pruning, nguyen2023lsr}, after which the resulting token-weight dictionaries are indexed using Lucene’s impact index. To reproduce SPARTA, we follow the indexing procedure described in the original publication, computing vocabulary-wide term scores from contextualized embeddings, and selecting the top-weighted terms for indexing \citep{zhao2021sparta}.

\subsubsection{Dense and Expansion-Based Retrievers}
\label{sec:dense-ret}
We implement dense and expansion-based retrievers using the \texttt{transformers} package \citep{wolf2020transformers} and index with \texttt{faiss-cpu} \citep{douze2024faiss}. For ColBERTv2, we use the official ColBERT GitHub repository\footnote{\url{https://github.com/stanford-futuredata/ColBERT}} and the same hyperparameters as the authors of the ColBERTv2 paper \citep{santhanam2022colbertv2}. Unlike the original setup, we fix the number of $k$-means iterations to $k=4$ across all datasets to keep the runtime of the clustering stage manageable (as we discuss in Section~\ref{sec:results}, this does not substantially affect the results). For prompt templates and configuration choices (e.g., context length), we follow recommendations from model papers, Hugging Face model cards, and GitHub repositories, with minor adaptations for instruction-following models. For docT5query, we generate three queries per document and perform retrieval using BM25 with default parameters. For query2doc, we use \texttt{Qwen2.5-7B-Instruct} \citep{qwenteam2025qwen2.5} to generate pseudo-documents, as the API-based generation model used in the original publication (GPT-4) is no longer available. Analogous to query2doc, for HyDE we use \texttt{Qwen2.5-7B-Instruct} \citep{qwenteam2025qwen2.5} to generate hypothetical documents, as the API-based generation model used in the original publication (InstructGPT) is no longer available. In addition, for RepLLaMA we use a context length of 1,024 (vs. 2,048), which balances retrieval effectiveness and runtime \citep{ma2024repllama}. For the instruction-based retrievers \textsc{InstructOR} (large), LLM2Vec, \textsc{GritLM}, NV-Embed-v2 and KaLM-Embedding (v2.5), we use a single unified prompt for all datasets that is based on their task-specific prompts.

\subsubsection{Comparability and Implementation Variability}

Our inclusion criteria in Section \ref{sec:retrievers} yield a heterogeneous set of retrievers, ranging from traditional lexical matching approaches to instruction-following general-purpose embedding models. In addition to architectural and training differences, our analysis of the corresponding literature revealed substantial variability in evaluation implementations and software infrastructure. Existing studies employ different retrieval frameworks, indexing backends, preprocessing pipelines, and evaluation toolkits, often relying on custom implementations that prohibit direct comparability between reported results. For example, of the method papers in Table \ref{tab:retrievers}, 45\% reported using standardized frameworks such as Anserini/Pyserini, BEIR, or MTEB, 30\% relied on custom evaluation pipelines, 15\% employed a mix, and 10\% did not state their approach. Furthermore, 67\% of the method papers reuse results from prior work, typically obtained under different experimental setups, which also affects comparability. Although efforts have been made to provide unified evaluation frameworks for retrieval models, such as BEIR \citep{thakur2021beir}, MTEB \citep{muennighoff2023mteb}, Anserini/Pyserini \citep{yang2018anserini, lin2021pyserini}, and PyTerrier \citep{macdonald2021pyterrier}, remain limited both in terms of supported retrievers and dataset coverage. Consequently, the goal of our experimental setup is to ensure the fairest possible comparison between retrievers by enforcing near-identical evaluation conditions across all models, which motivates the design choices described in Sections \ref{sec:sparse-ret} and \ref{sec:dense-ret}.

\subsubsection{Hardware}
We conducted all our experiments on a compute cluster consisting of GPU nodes equipped with four NVIDIA H100 GPUs (94\,GB VRAM), and two AMD EPYC 9454 processors, each with 48 cores, and 768\,GB of system RAM.

\section{Results}
\label{sec:results}

We break down our findings into retrieval performance, runtimes, and an investigation of model similarities based on overlaps between the subsets of queries that the retrievers struggle to answer.

\begin{table*}[ht]
    \centering
    \footnotesize
    \caption[Evaluation results --- all retrievers on all datasets.]{Evaluation results --- all retrievers on all datasets. S$@5$ = Success$@5$ and R$@10$ = Robustness$@10$. Score is the average across MAP$@1000$ and nDCG$@5$. \textcolor{Black}{Black} = new result, \textcolor{SpecBlue}{blue} = within 2 standard errors of the published result, \textcolor{SpecTeal}{teal} = between 2 and 3 standard errors above the published result, \textcolor{ForestGreen}{green} = 3 or more standard errors above the published result, \textcolor{SpecPurple}{purple} = between 2 and 3 standard errors below the published result, \textcolor{SpecRed}{red} = 3 or more standard errors below the published result. The best result per dataset is highlighted bold.}
    \begin{tabular}{lccccccccc}
        \toprule
        \textbf{Dataset} $\rightarrow$ & \textbf{MS MARCO} & \textbf{TREC-DL 19} & \textbf{TREC-DL 20} & \textbf{BEIR} & \textbf{LoTTE} & \multicolumn{2}{c}{\textbf{InstructIR}} & \multicolumn{2}{c}{\textbf{FollowIR}} \\
        \cmidrule(lr){2-2} \cmidrule(lr){3-3} \cmidrule(lr){4-4} \cmidrule(lr){5-5} \cmidrule(lr){6-6} \cmidrule(lr){7-8} \cmidrule(lr){9-10}
        \textbf{Model} $\downarrow$ & \textbf{MRR$@10$} & \textbf{nDCG$@10$} & \textbf{nDCG$@10$} & \textbf{nDCG$@10$} & \textbf{S$@5$} & \textbf{R$@10$} & \textbf{nDCG$@10$} & \textbf{Score} & \textbf{$p$-MRR} \\
        \midrule
        \multicolumn{10}{l}{\textbf{Sparse Retrievers}}\\
        \midrule
        \addlinespace[2pt]
        \multicolumn{9}{l}{\textit{Lexical}} \\
        \addlinespace[1pt]
        BM25 \citep{robertson2009bm25} & \textcolor{SpecBlue}{0.185} & \textcolor{SpecBlue}{0.512} & \textcolor{SpecBlue}{0.477} & \textcolor{SpecBlue}{0.427} & \textcolor{SpecBlue}{0.477} & \textcolor{SpecBlue}{0.277} & \textcolor{SpecBlue}{0.766} & \textcolor{SpecBlue}{16.680} & \textcolor{SpecBlue}{-2.324} \\
        \addlinespace[2pt]
        \multicolumn{9}{l}{\textit{Learned Sparse}} \\
        \addlinespace[1pt]
        DeepCT \citep{dai2020deepct} & \textcolor{SpecRed}{0.180} & 0.507 & 0.447 & \textcolor{SpecGreen}{0.375} & 0.389 & 0.282 & 0.772 & 13.227 & -2.440 \\
        SPARTA \citep{zhao2021sparta} & 0.142 & 0.351 & 0.285 & \textcolor{SpecRed}{0.274} & 0.344 & 0.076 & 0.542 & 7.738 & \textbf{6.257} \\
        uniCOIL \citep{lin2021unicoil} & \textcolor{SpecBlue}{0.307} & \textcolor{SpecBlue}{0.648} & 0.642 & \textcolor{SpecBlue}{0.446} & 0.577 & 0.284 & 0.731 & 19.599 & 0.182 \\
        SPLADE-v3 \citep{lassance2024spladev3} & \textbf{\textcolor{SpecBlue}{0.403}} & \textcolor{SpecBlue}{0.725} & \textcolor{SpecBlue}{0.752} & \textcolor{SpecPurple}{0.508} & \textcolor{SpecBlue}{0.701} & 0.498 & 0.873 & 20.559 & -2.794 \\
        \midrule
        \multicolumn{10}{l}{\textbf{Dense Retrievers}}\\
        \midrule
        \addlinespace[2pt]
        \multicolumn{9}{l}{\textit{Encoder-Only}} \\
        \addlinespace[1pt]
        DPR \citep{karpukhin2020dpr} & 0.106 & 0.297 & 0.270 & \textcolor{SpecBlue}{0.251} & 0.279 & 0.211 & 0.637 & 11.630 & 1.167 \\
        ColBERTv2 \citep{santhanam2022colbertv2} & \textcolor{SpecBlue}{0.397} & \textcolor{SpecBlue}{0.746} & \textcolor{SpecBlue}{0.750} & \textcolor{SpecPurple}{0.490} & \textcolor{SpecBlue}{0.675} & \textcolor{SpecBlue}{0.142} & \textcolor{SpecBlue}{0.686} & 17.508 & 0.051 \\
        TAS-B \citep{hofstatter2021tasb} & \textcolor{SpecBlue}{0.340} & \textcolor{SpecBlue}{0.704} & \textcolor{SpecBlue}{0.688} & \textcolor{SpecBlue}{0.427} & 0.572 & 0.428 & 0.825 & 18.447 & -3.210 \\
        ANCE \citep{xiong2021ance} & \textcolor{SpecRed}{0.275} & \textcolor{SpecBlue}{0.592} & 0.590 & \textcolor{SpecBlue}{0.399} & \textcolor{SpecRed}{0.567} & 0.325 & 0.718 & 14.562 & -2.750 \\
        SimCSE \citep{gao2021simcse} & 0.067 & 0.251 & 0.190 & \textcolor{SpecRed}{0.244} & 0.270 & 0.163 & 0.651 & 12.229 & 0.860 \\
        RetroMAE \citep{xiao2022retromae} & \textcolor{SpecPurple}{0.344} & \textcolor{SpecBlue}{0.679} & \textcolor{SpecBlue}{0.724} & \textcolor{SpecBlue}{0.469} & \textcolor{SpecBlue}{0.622} & 0.435 & 0.835 & 17.927 & -3.054 \\
        coCondenser \citep{gao2022cocondenser} & \textcolor{SpecRed}{0.340} & \textcolor{SpecBlue}{0.685} & 0.674 & \textcolor{SpecRed}{0.399} & \textcolor{SpecBlue}{0.564} & 0.367 & 0.744 & 13.726 & -1.231 \\
        Contriever \citep{izacard2022contriever} & \textcolor{SpecBlue}{0.337} & \textcolor{SpecBlue}{0.670} & \textcolor{SpecBlue}{0.675} & \textcolor{SpecBlue}{0.464} & \textcolor{SpecBlue}{0.621} & \textcolor{SpecBlue}{0.474} & \textcolor{SpecBlue}{0.848} & \textcolor{SpecBlue}{19.074} & \textcolor{SpecBlue}{-2.216} \\
        COCO-DR (large) \citep{yu2022cocodr} & \textcolor{SpecBlue}{0.349} & \textcolor{SpecBlue}{0.729} & \textcolor{SpecBlue}{0.698} & \textcolor{SpecRed}{0.489} & \textcolor{SpecGreen}{0.641} & 0.428 & 0.809 & 17.393 & -1.108 \\
        \textsc{SimLM} \citep{wang2023simlm} & \textcolor{SpecRed}{0.342} & \textcolor{SpecBlue}{0.650} & \textcolor{SpecBlue}{0.640} & 0.335 & 0.462 & 0.271 & 0.712 & 5.165 & -4.443 \\
        \textsc{Dragon}+ \citep{lin2023dragon} & \textcolor{SpecBlue}{0.382} & \textcolor{SpecBlue}{0.736} & \textcolor{SpecBlue}{0.720} & \textcolor{SpecBlue}{0.488} & \textcolor{SpecBlue}{0.674} & 0.498 & 0.869 & 20.491 & -1.734 \\
        mE5 (large) \citep{wang2024me5} & \textcolor{SpecRed}{0.300} & 0.633 & 0.673 & \textcolor{SpecGreen}{0.533} & 0.752 & 0.570 & 0.893 & 22.146 & 0.154 \\
        GTE (large v1.5) \citep{li2023gte} & \textcolor{SpecBlue}{0.361} & 0.708 & 0.729 & \textcolor{SpecBlue}{0.587} & 0.794 & 0.582 & 0.892 & 21.724 & -0.085 \\
        BGE (large v1.5) \citep{xiao2024bge} & \textcolor{SpecBlue}{0.353} & 0.697 & 0.698 & \textcolor{SpecRed}{0.528} & 0.743 & 0.521 & 0.869 & \textcolor{SpecBlue}{17.369} & \textcolor{SpecBlue}{-1.838} \\
        Jina Embeddings v3 \citep{sturua2025jina3} & \textcolor{SpecBlue}{0.341} & 0.699 & 0.683 & \textcolor{SpecRed}{0.535} & 0.784 & 0.488 & 0.835 & 23.944 & 0.519 \\
        Nomic Embed (v1.5) \citep{nussbaum2025nomic} & \textcolor{SpecBlue}{0.360} & 0.716 & 0.701 & \textcolor{SpecBlue}{0.536} & 0.716 & 0.429 & 0.801 & 19.634 & -1.879 \\
        \addlinespace[2pt]
        \multicolumn{9}{l}{\textit{Encoder-Decoder}} \\
        \addlinespace[1pt]
        GTR (large) \citep{ni2022gtr} & \textcolor{SpecPurple}{0.368} & 0.713 & 0.687 & \textcolor{SpecBlue}{0.476} & 0.761 & \textcolor{SpecBlue}{0.376} & \textcolor{SpecBlue}{0.760} & 17.868 & -5.320 \\
        \textsc{InstructOR} (large) \citep{su2023instructor} & 0.343 & 0.645 & 0.668 & \textcolor{SpecPurple}{0.471} & 0.768 & \textcolor{SpecBlue}{0.562} & \textcolor{SpecRed}{0.887} & 20.332 & -3.138 \\
        \addlinespace[2pt]
        \multicolumn{9}{l}{\textit{Decoder-Only}} \\
        \addlinespace[1pt]
        LLM2Vec \citep{behnamghader2024llmvec} & 0.317 & 0.697 & 0.685 & 0.523 & 0.797 & 0.639 & 0.916 & 25.237 & 5.540 \\
        RepLLaMA \citep{ma2024repllama} & \textcolor{SpecPurple}{0.399} & \textcolor{SpecBlue}{0.741} & \textcolor{SpecBlue}{0.728} & \textcolor{SpecRed}{0.534} & 0.762 & \textcolor{SpecRed}{0.504} & \textcolor{SpecRed}{0.864} & \textcolor{SpecBlue}{21.382} & \textcolor{SpecBlue}{-2.299} \\
        \textsc{GritLM} \citep{muennighoff2025gritlm} & 0.356 & 0.721 & 0.730 & \textcolor{SpecRed}{0.571} & 0.853 & \textbf{0.753} & \textbf{0.951} & \textbf{\textcolor{SpecBlue}{27.564}} & \textcolor{SpecTeal}{5.537} \\
        Qwen3 Embedding (0.6B) \citep{zhang2025qwen3embedding} & 0.309 & 0.665 & 0.661 & 0.529 & 0.689 & 0.460 & 0.817 & 24.264 & 5.363 \\
        NV-Embed-v2 \citep{lee2025nvembed} & 0.389 & \textbf{0.753} & \textbf{0.765} & \textbf{\textcolor{SpecRed}{0.612}} & \textbf{0.876} & 0.677 & 0.928 & 26.800 & -0.355 \\
        EmbeddingGemma \citep{vera2025embeddinggemma} & 0.317 & 0.718 & 0.714 & 0.560 & 0.774 & 0.602 & 0.906 & 23.832 & 1.043 \\
        KaLM-Embedding (v2.5) \citep{zhao2026kalmembeddingv2} & 0.265 & 0.568 & 0.582 & \textcolor{SpecRed}{0.470} & 0.645 & 0.553 & 0.873 & 22.691 & -1.508 \\
        \midrule
        \multicolumn{10}{l}{\textbf{Expansion-Based Retrievers}}\\
        \midrule
        docT5query \citep{nogueira2019doct5query} & \textcolor{SpecRed}{0.246} & \textcolor{SpecBlue}{0.602} & \textcolor{SpecBlue}{0.576} & \textcolor{SpecBlue}{0.443} & 0.542 & 0.349 & 0.799 & 16.820 & -2.245 \\
        query2doc \citep{wang2023query2doc} & \textcolor{SpecPurple}{0.204} & \textcolor{SpecBlue}{0.634} & \textcolor{SpecBlue}{0.598} & 0.462 & 0.495 & 0.356 & 0.803 & 19.712 & -2.614 \\
        HyDE \citep{gao2023hyde} & 0.302 & \textcolor{SpecGreen}{0.728} & \textcolor{SpecGreen}{0.716} & 0.498 & 0.658 & 0.464 & 0.836 & 21.322 & -0.744 \\
        \bottomrule
    \end{tabular}
    \label{tab:retrieval-quality}
\end{table*}

\subsection{Retrieval Effectiveness}

In Table~\ref{tab:retrieval-quality}, we show the retrieval effectiveness of all tested retrievers on all datasets, using their default metrics. Additional disaggregated results are available in Tables \ref{tab-a:retrieval-quality-zscores}--\ref{tab-a:instructir-followir-results}.

\subsubsection{Overall Performance} 

Considering overall performance, we find that dense decoder-only models lead the pack, with \textbf{NV-Embed-v2} performing best on both TREC-DL datasets, BEIR, and LoTTE, and \textbf{\textsc{GritLM}} performing best on \textsc{InstructIR} and \textsc{FollowIR}. The exceptions are \textbf{SPLADE-v3}, which has the best performance on MS MARCO, and \textbf{SPARTA} on $p$-MRR on \textsc{FollowIR}, both of which are sparse retrievers.
Notable performances are \textbf{SPLADE-v3} and \textbf{\textsc{Dragon}+} on the TREC-DL datasets. Similarly noteworthy is the overall performance of \textbf{GTE (large v1.5)}, which outperforms substantially larger retrievers on the classic datasets.

When comparing the retriever families, we find generalizable trends. Dense retrievers demonstrate strong performance across all datasets, outperforming sparse and expansion-based retrievers on LoTTE and \textsc{InstructIR} by the largest margins.
In particular, encoder- and decoder-only retrievers are best across the classic datasets (MS MARCO, BEIR, and LoTTE), although all are outperformed by the sparse SPLADE-v3 on MS MARCO. Unsurprisingly, retrievers with instruction-tuning perform best on the instruction-following datasets (\textsc{InstructIR} and \textsc{FollowIR}).

Finally, we make a few other observations of general trends. 
First, we find models that rely on textual prompts to be (extremely) sensitive to the chosen prompt. While some authors go as far as using dataset-specific prompts (which generally increases performance), no guidelines for prompt engineering are provided, thereby likely obfuscating a major source of performance gain.
Second, for the general embedding models, we find that, as reported in the papers, in some cases the older English-only versions outperform the newer multilingual versions (e.g., for GTE vs. mGTE and BGE vs. BGE-M3), while in other cases it is the other way around (e.g., mE5 vs. E5).

\subsubsection{Comparison to Published Results}
\label{subsec:deviations}
To assess the reproducibility of previously published results within the selected common evaluation framework, we compiled the reported performances of all tested models from the respective publications (where available) and compared them to the results from our experiments. For each model-dataset pair with a corresponding literature value, we compute a standardized deviation $\Delta_\sigma$ as
\begin{equation}
\Delta_\sigma = \frac{\hat{\mu} - m_{\mathrm{pub}}}{\widehat{\mathrm{SE}}_{\mathrm{boot}}}
\end{equation}
where $m_\text{pub}$ denotes the published score, $\hat\mu$ denotes the mean of our results and $\widehat{\text{SE}}_\text{boot}$ the standard error of our results that we compute using bootstrap resampling \citep{efron1986bootstrap} from $B=1000$ samples. Thus, $\Delta_\sigma$ measures how far the published result lies from our reproduced score in units of bootstrap-estimated standard errors. We emphasize that \textit{a large deviation is not necessarily an implementation or reporting error in the literature}, but more likely indicates a strong dependence on implementation details that may render such results in the literature incomparable as we discuss in the following.

The results of this comparison are shown in Table~\ref{tab:retrieval-quality}, where results \textcolor{SpecBlue}{close to the literature values} ($|\Delta_\sigma| \le 2$) are highlighted in blue. Deviations between two and three standard errors are indicated as \textcolor{SpecTeal}{moderately higher} ($2 < \Delta_\sigma < 3$) or \textcolor{SpecPurple}{moderately lower} ($-3 < \Delta_\sigma < -2$), while deviations of at least three standard errors are marked as \textcolor{ForestGreen}{considerably higher} ($\Delta_\sigma \ge 3$) or \textcolor{SpecRed}{considerably lower} ($\Delta_\sigma \le -3$) relative to the published results. Uncolored entries denote new, previously unpublished results.

Overall, we find that reproducibility is good, with only small deviations (\textcolor{SpecBlue}{blue results}) in most instances (76/109 or 69.7\% of the published results), yet also note that the majority of our results are not available in the literature (188/297 or 63.3\%).
Of the 33 results with large deviations (30.3\% of the 109 published results), 20 (60.6\%) are at least three standard deviations worse than the published results, while 5 (15.2\%) are at least three standard deviations better.
In the following, we review all results with large deviations ($|\Delta_{\sigma}| \ge 2$) and report the differences between our results and the published results ($\Delta = s_\text{our} - s_\text{pub}$), using the color scheme from Table \ref{tab:retrieval-quality}.

\textbf{DeepCT} performs worse in our experiments on MS MARCO ($\Delta=\textcolor{SpecRed}{-0.063}$) yet better on BEIR ($\Delta=\textcolor{SpecGreen}{+0.065}$).
We assume that this difference is due to a re-implementation of the model, for which only one checkpoint is available on Hugging Face that is different from the one used by the authors in the original publication~\citep{dai2020deepct}.
Furthermore, recent work has shown that learned sparse retrievers can be highly sensitive to preprocessing choices, such as casing, which may substantially affect retrieval effectiveness across datasets \citep[cf.][]{lionis2026casing}. The increased score on BEIR may be due to our usage of hyperparameters from the original publication instead of the default Anserini hyperparameters used by the BEIR authors \citep{thakur2021beir}.

\textbf{HyDE} obtained substantially higher scores on both TREC-DL 2019 ($\Delta=\textcolor{SpecGreen}{+0.115}$) and TREC-DL 2020 ($\Delta=\textcolor{SpecGreen}{+0.137}$), which we attribute to the fact that we had to use a more recent and advanced generative LLM to generate pseudo-documents (Qwen2.5-7B-Instruct instead of InstructGPT, which is no longer available).

\textbf{SPARTA} performed worse on BEIR ($\Delta=\textcolor{SpecRed}{-0.081}$).
Similar to DeepCT, due to a lack of availability, we had to partially re-implement this model from a Hugging Face checkpoint that differs from the one used in the original publication. Due to varying available specifications it is possible that our $k$ hyperparameter for top-$k$ sampling also varied from the original (both 512 and 2000 are cited). Furthermore, a recent study has shown that learned sparse retrieval models can be sensitive to preprocessing  choices, which may further contribute to discrepancies across implementations and evaluation settings \citep{lionis2026casing}.

The slightly lower scores of \textbf{ColBERTv2} on BEIR ($\Delta=\textcolor{SpecPurple}{-0.009}$) and \textbf{SPLADE-v3} on BEIR ($\Delta=\textcolor{SpecPurple}{-0.009}$) may stem from differences in hyperparameter choices. As noted in Section~\ref{sec:implementation}, we only used four iterations for the $k$-means clustering during ColBERTv2's index generation, compared to the 20 iterations reported in the original implementation\footnote{\url{https://github.com/stanford-futuredata/ColBERT}}. Additionally, we applied multiple post-processing steps for SPLADE-v3 due to the need to reimplement parts of the method, which may account for the observed deviations.

On the MS MARCO dataset, a group of dense models performed notably worse in our experiments, including \textbf{GTR (large)} ($\Delta=\textcolor{SpecPurple}{-0.011}$), \textbf{RetroMAE} ($\Delta=\textcolor{SpecPurple}{-0.010}$), \textbf{\textsc{SimLM}} ($\Delta=\textcolor{SpecRed}{-0.069}$), \textbf{ANCE} ($\Delta=\textcolor{SpecRed}{-0.055}$), and \textbf{coCondenser} ($\Delta=\textcolor{SpecRed}{-0.042}$).
One possible explanation is that the original results were obtained using a version of MS MARCO that includes documents titles, which tends to increase performance across the board \citep{lassance2023msmarco}. 
For RetroMAE, we used a model checkpoint fine-tuned on the BEIR datasets for all experiments. In the original publication \citep{xiao2022retromae}, however, the reported MS MARCO results appear to have been obtained using a checkpoint fine-tuned on MS MARCO\footnote{\url{https://huggingface.co/Shitao/RetroMAE_MSMARCO_finetune}}. This difference in training data may partly explain the observed performance gap.
For ANCE and \textsc{SimLM}, it is unclear if we may also have used a different model checkpoint due to availability and sparse documentation.

A second group of models shows slightly lower scores than those reported in the literature. This includes
\textbf{coCondenser} on BEIR ($\Delta=\textcolor{SpecRed}{-0.014}$), 
\textbf{COCO-DR} on BEIR ($\Delta=\textcolor{SpecRed}{-0.016}$), 
\textbf{\textsc{InstructOR}} on BEIR ($\Delta=\textcolor{SpecPurple}{-0.010}$) and \textsc{InstructIR} (nDCG$@10$ $\Delta=\textcolor{SpecRed}{-0.008}$), 
\textbf{SimCSE} on BEIR ($\Delta=\textcolor{SpecRed}{-0.013}$), 
\textbf{Jina Embeddings v3} on BEIR ($\Delta=\textcolor{SpecRed}{-0.013}$), and \textbf{\textsc{GritLM}} on BEIR ($\Delta=\textcolor{SpecRed}{-0.014}$). 
These discrepancies may be attributable to implementation or evaluation differences, such as the handling of document titles in BEIR, as well as sampling variation introduced by the bootstrapped mean. Overall, however, the magnitude of the deviations ($\Delta$) remains small.

Similarly, a group of retrievers with instruction-following capabilities performed worse in our experiments, including 
\textbf{BGE (large v1.5)} on BEIR ($\Delta=\textcolor{SpecRed}{-0.023}$), 
\textbf{NV-Embed-v2} on BEIR ($\Delta=\textcolor{SpecRed}{-0.027}$), 
\textbf{mE5 (large)} on MS MARCO ($\Delta=\textcolor{SpecRed}{-0.040}$), 
and \textbf{KaLM-Embedding (v2.5)} on BEIR ($\Delta=\textcolor{SpecRed}{-0.09}$). 
We suspect that prompt variations are the reason for these performance differences, as these methods originally used prompts that were tailored to the datasets. Specifically KaLM-Embedding (v2.5) is seemingly very sensitive to the wording of the instruction in the query.
In contrast, the performance of \textbf{\textsc{GritLM}} increases on \textsc{FollowIR} ($p$-MRR $\Delta=\textcolor{SpecTeal}{+5.53}$) and \textbf{mE5 (large)} performs better on BEIR ($\Delta=\textcolor{SpecGreen}{+0.013}$), which we also attribute to prompt differences.

For \textbf{docT5query}, we observe a slightly worse performance on MS MARCO ($\Delta=\textcolor{SpecRed}{-0.013}$).
A possible explanation is the fact that we generated fewer queries (3 vs.\ the 20 in the original) to keep execution times at a feasible level, following the intuition provided by \citet{gospodinov2023doc2query--} that fewer queries may be sufficient. This could also explain the worse performances on TREC-DL 19 ($\Delta=\textcolor{SpecBlue}{-0.040}$) and TREC-DL 20 ($\Delta=\textcolor{SpecBlue}{-0.043}$), as we reused the embeddings from MS MARCO for the TREC-DL datasets.

\textbf{query2doc} performed slightly worse on MS MARCO ($\Delta=\textcolor{SpecPurple}{-0.010}$).
A likely explanation is our use of a less powerful LLM for generating the pseudo-documents (\texttt{Qwen2.5-7B-Instruct} instead of GPT-4). In contrast to the original implementation, we also consistently used zero-shot prompting instead of $k$-shot prompting for generating the pseudo-documents, since not all datasets we evaluated on contain examples that can be used for this purpose.

\textbf{RepLLaMA} performs worse on multiple datasets: MS MARCO ($\Delta=\textcolor{SpecPurple}{-0.013}$), BEIR ($\Delta=\textcolor{SpecRed}{-0.017}$), and \textsc{InstructIR} (Robustness$@10$ $\Delta=\textcolor{SpecRed}{-0.022}$, nDCG$@10$ $\Delta=\textcolor{SpecRed}{-0.012}$). A possible explanation is that we used a shorter context length than the authors (1024 vs. 2048) to significantly reduce runtime to feasible levels, while staying close to the originally published results \citep{ma2024repllama}.

Finally, we observe two results that deviate from the published scores on LoTTE, namely \textbf{ANCE} ($\Delta=\textcolor{SpecRed}{-0.044}$) and \textbf{COCO-DR} ($\Delta=\textcolor{SpecGreen}{+0.020}$).
In both cases, despite our best efforts, we do not have a satisfactory explanation for the differences in performance.

The methods reproducing most consistently across all datasets are 
\textbf{ColBERTv2} ($|\bar\Delta|=0.0019$),
\textbf{Nomic Embed (v1.5)} ($|\bar\Delta|=0.0026$), 
\textbf{SPLADE-v3} ($|\bar\Delta|=0.0033$),
and
\textbf{GTE (large v1.5)} ($|\bar\Delta|=0.0036$), 
where $|\bar\Delta| = \frac{1}{N} \sum^N_{i=1} |\Delta_i|$.
Notably, of the ten smallest mean differences, nine are dense retrievers.
In addition, 
\textbf{BM25}, \textbf{uniCOIL}, \textbf{TAS-B}, \textbf{Contriever}, and \textbf{\textsc{Dragon}+} show exclusively non-significant differences (all \textcolor{SpecBlue}{blue results}), while \textbf{SPLADE-v3}, \textbf{ColBERTv2}, \textbf{RetroMAE}, and \textbf{GTR (large)} exhibit only one moderate deviation (\textcolor{SpecPurple}{purple result}).

Aggregated by dataset across models, the largest deviations occur on 
MS MARCO ($|\bar\Delta|=0.017$), 
TREC-DL 19 ($|\bar\Delta|=0.025$), 
and TREC-DL 20 ($|\bar\Delta|=0.024$), 
which we attribute largely to the inconsistent handling of document titles in the literature \citep{lassance2023msmarco}.

\subsection{Query Runtime Analysis}
To consider the performance-runtime trade-off, we collect the average end-to-end query latency in milliseconds for each model-dataset pair. Following standard practice \citep{lassance2022efficiency, santhanam2023moving}, the average query latency $L_{ms}$ is the sum of the total time spent in query encoding $t_{enc}$ and retrieval/search $t_{ret}$, normalized by the number of queries $N$:
\begin{equation}
    L_{ms} = \frac{t_{enc} + t_{ret}}
    {N}. 
\end{equation}
Notably, we report average runtimes by amortizing the loading times of indices across queries since the variety of implementation strategies among models renders their exclusion infeasible.

\begin{figure}
  \hspace*{-4mm} 
  \includegraphics[width = 0.49\textwidth]{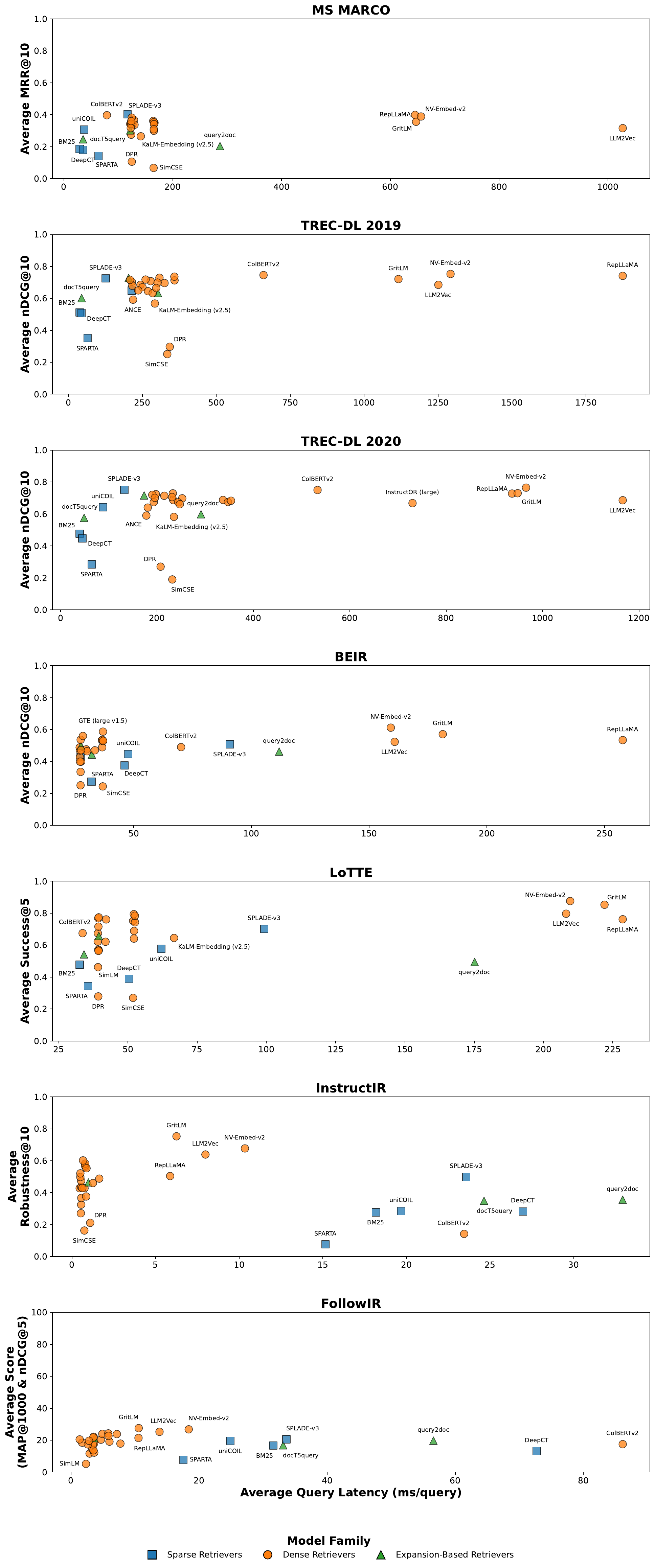}
  \caption{Retrieval performance vs.\ average query times on all datasets. Selected outliers are labeled.}
  \label{fig:runtimes-retquality}
\end{figure}

\subsubsection{Trends and Outliers}
To highlight general performance trends and outliers, we provide a visualization of retrieval performances vs.\ average query times in Figure~\ref{fig:runtimes-retquality}.
Overall, we find that retriever families follow the expected trends, with sparse retrievers being the fastest (44 ms/query), followed by expansion-based retrievers (68 ms/query), and dense models (75 ms/query). Particularly the decoder-only dense retrievers have high query latencies (154 ms/query). This highlights a steep cost for the performance of state-of-the-art models over more lightweight approaches that does not seem to be universally warranted \citep{abdallah2026llmbased}. 
Notably, however, sparse retrievers are $3-5\times$ slower than the other families on \textsc{InstructIR} and \textsc{FollowIR}, where dense retrievers without an LLM-backbone are very fast. We attribute this to the ratio of few queries to the much larger size of the document collections, resulting in a pronounced index-induced overhead. 
A notable outlier is ColBERTv2, a dense retriever, which is quite fast on MS MARCO and LoTTE, yet exceedingly slow on all other datasets.

\subsubsection{Breakdown by Dataset}
Unsurprisingly, we find the fastest model to be BM25 (28 ms/query), while the slowest model is RepLLaMA (229 ms/query). In the following, we highlight the performances of notable retrievers on each of the datasets.

\textbf{MS MARCO.}
We find that SPLADE-v3 performs best (MRR@10 of 0.403) with a slightly lower than average query time (117 ms/query). ColBERTv2 has the 3rd best performance (MRR@10 of 0.397) with a very low query time (79 ms/query).

\textbf{TREC-DL 2019 and 2020.}
Here, NV-Embed-v2 performs best (nDCG@10 of 0.753) with a very high query time (1,292 ms/query), while ColBERTv2 has a comparable performance (nDCG@10 of 0.746) at half the average query time (659 ms/query). The performance of \textsc{Dragon}+ is similarly close (nDCG@10 of 0.736) with an even lower query time (358 ms/query), while SPLADE-v3 provides likely the optimal trade-off with its 6th best performance (nDCG@10 of 0.725) and a query time of just 126 ms/query. The result of \textbf{TREC-DL 2020} is very similar (NV-Embed-v2 in first and SPLADE-v3 in second).

\textbf{BEIR.}
The top performance of NV-Embed-v2 (nDCG@10 of 0.612) comes at the cost of just a moderate query time (159 ms/query), while GTE (large v1.5) is substantially faster (37 ms/query) for the 2nd best performance (nDCG@10 of 0.587).

\textbf{LoTTE.}
NV-Embed-v2 again has the best performance (Success@5 of 0.876) at a relatively high query time (210 ms/query), while GTE (large v1.5)'s performance is still close (Success@5 of 0.794) at just 52 ms/query.

\textbf{\textsc{InstructIR}.}
Here, \textsc{GritLM} provides the best performance (Robustness@10 of 0.753) with an acceptable runtime (6.3 ms/query), while EmbeddingGemma has the 4th best performance (Robustness@10 of 0.602) at a very low query time (0.66 ms/query).

\textbf{\textsc{FollowIR}.}
\textsc{GritLM} also performs best on this dataset (Score of 27.564) with an acceptable runtime (10.6 ms/query), while Qwen3 Embedding (0.6B) is substantially faster (5.8 ms/query) for a comparable performance (Score of 24.264).

\subsubsection{The Cost of the Last Increment}
On average across all datasets the fastest model is BM25 ($28$ ms/query) and the slowest RepLLaMA ($229$ ms/query), but the more informative quantity is what the final increment of effectiveness costs on each dataset. On LoTTE, moving from GTE (large v1.5) ($52$ ms, Success$@5$ of $0.794$) to NV-Embed-v2 ($210$ ms, $0.876$) costs $158$ ms for a gain of $0.082$, or $1.9$ ms per $0.001$. On BEIR for the same two models, the cost is $122$ ms for $0.025$ nDCG$@10$ ($37$ vs. $159$ ms), or $4.9$ ms per $0.001$. On TREC-DL 2019, by contrast, SPLADE-v3 reaches $0.725$ at $126$ ms/query while NV-Embed-v2 reaches $0.753$ at $1{,}292$ ms: $42$ ms per $0.001$, more than twenty times the LoTTE rate. The instruction-following datasets invert the pattern entirely, with \textsc{GritLM} both the most effective and among the fastest (Robustness$@10$ of $0.753$ at $6.3$ ms/query; Score of $27.564$ at $10.6$ ms/query). Thus, how the best model performs relative to cheaper alternatives is a property of the dataset, not a fixed property of the model.

\subsection{Failure Point Analysis}
\label{sec:failure-point-ana}

\begin{figure}
  \includegraphics[width = 0.495\textwidth]{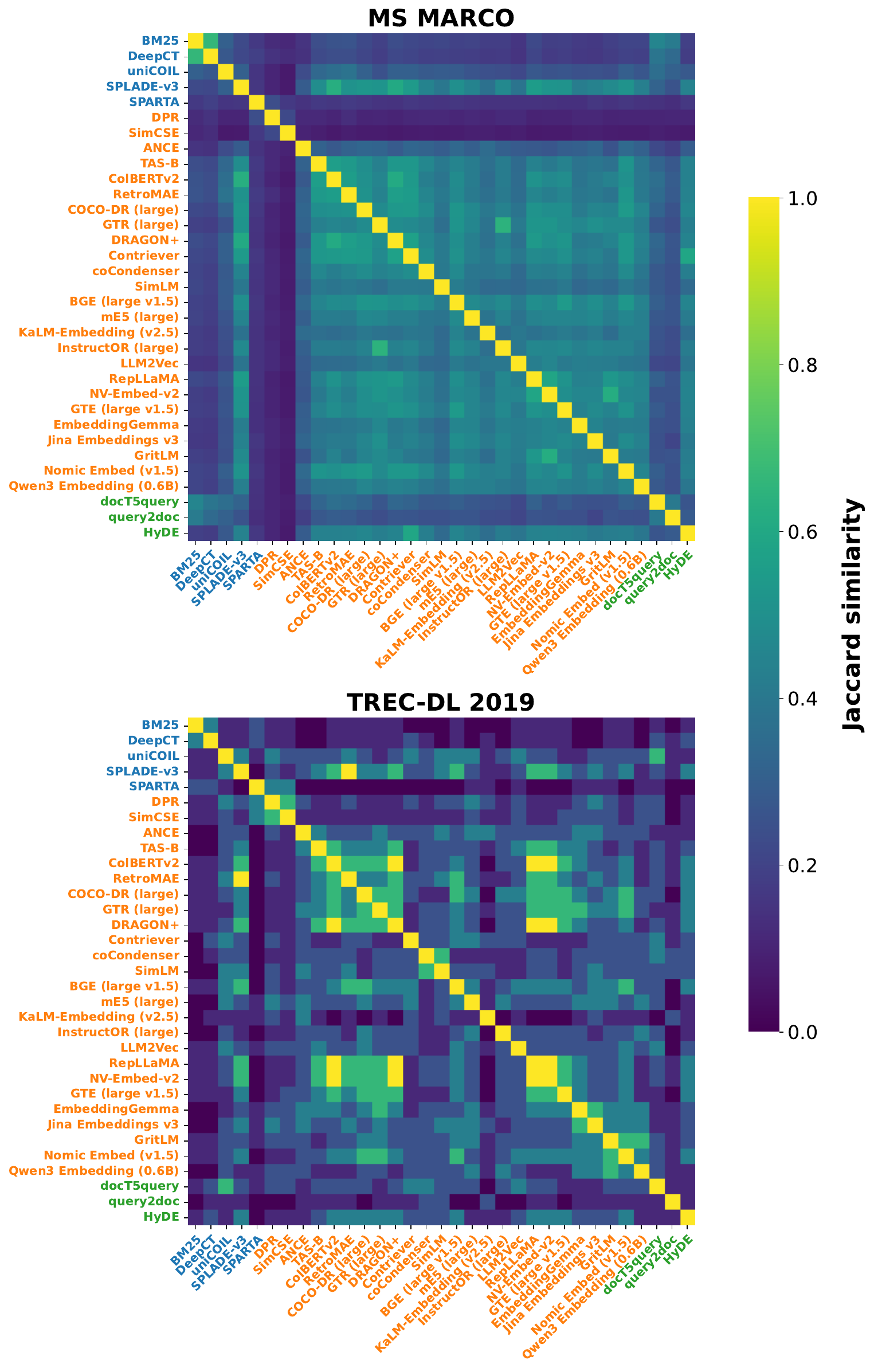}
  \caption{Jaccard similarities of hard-query sets. Pairwise Jaccard similarity between models based on the overlap of their hard-query sets for MS MARCO (top) and TREC-DL 2019 (bottom), using the hardest 10\% of queries per dataset.}
  \label{fig:hard-queries}
\end{figure}

To investigate similarities between the failure points of the retrievers, we conduct a comparison of queries on which they perform poorly. To this end, we use the rank of the highest-ranked relevant document in a model’s list of retrieved documents as an indicator of query performance on a given dataset, capped at $k=1000$ (i.e., if no relevant document is retrieved within depth 1000, the query is assigned a rank of 1001).
Based on this ranking, we then define the \textit{worst-performing queries} as the bottom $10\%$ of queries per model-dataset combination when sorting by rank in ascending order, thus guaranteeing a fixed number of queries per dataset-model pair. To measure the overlap between the worst-performing queries sets of two models on the same dataset, we compute the Jaccard similarity \citep{jaccard1902dist} and show the main results in Figure \ref{fig:hard-queries} and the supplementary results in Figure \ref{fig-a:hard-queries}.

Worst-performing query overlap analysis reveals both expected and unexpected patterns across retrievers and datasets. On \textbf{MS MARCO}, we observe moderate to high overlap between the worst-performing query sets of BM25 and DeepCT ($J \approx 0.6$), which is consistent with the fact that DeepCT relies on BM25 during retrieval. In contrast, SPLADE-v3 acts as an outlier, exhibiting relatively high similarity with the worst-performing query sets of most dense retrievers. Another notable pattern is a cross-shaped pattern formed by the three worst-performing retrievers (SPARTA, DPR, and SimCSE), which exhibit very low overlap with most other models ($J \approx 0.1$), indicating that they fail on substantially different queries. At the same time, the dense, instruction-tuned, and general embedding models show moderately high overlap among themselves ($J \approx 0.5$), suggesting that these approaches tend to struggle on somewhat similar subsets of queries.

Compared to this baseline, the \textbf{TREC-DL 2019} heatmap shows stronger clustering and more pronounced outliers, indicating greater variation in model behavior. Several high-similarity clusters emerge, including groups of dense retrievers, RepLLaMA-NV-Embed-v2-GTE (large v1.5), and pairs such as Contriever-coCondenser and EmbeddingGemma-Jina Embeddings v3. Interestingly, some clusters span models with substantially different backbones, suggesting that hard queries are not determined solely by architectural similarity. Finally, query expansion methods (docT5query, query2doc, and HyDE) struggle on largely distinct queries from their underlying retrievers (BM25 and Contriever), indicating that expansion shifts which queries are difficult for the retrieval method.

\section{Discussion}
In the following, we discuss our findings, focusing on (1) suggestions for retriever selection, (2) our experiences in replicating reported results, and (3) implications for further research.

\textbf{Performance.} We find that no single retriever consistently outperforms all others, and which retrievers do well varies with the benchmark. On MS MARCO and TREC-DL, strong performance is not tied to one family: the top retrievers include a learned sparse model (SPLADE-v3) alongside ColBERTv2, RepLLaMA and NV-Embed-v2. On the remaining datasets, the best results come consistently from more recently released models regardless of family. Robustness across datasets is therefore a more informative selection criterion than peak performance on any single benchmark.
 
\textbf{Recommendations.} Five take-aways follow from our results:

\begin{itemize}
    \item \textit{Size predicts latency, not effectiveness.} We suggest GTE as a general-purpose default balancing effectiveness and runtime, and NV-Embed-v2 only where latency is unconstrained.
    \item \textit{Most learned sparse methods are underperforming.} SPLADE-v3 is the only learned sparse method we recommend.
    \item \textit{Instruction-following is inconsistent and cannot be inferred from general performance.} 22 of 33 retrievers score a negative $p$-MRR. NV-Embed-v2 leads four datasets yet scores $-0.355$, while GritLM offers more balance (good all-rounder and very good on the instruction-following datasets). Prompt sensitivity compounds this, with KaLM-Embedding (v2.5) varying by $0.09$ nDCG@10 on BEIR primarily from wording, making prompt robustness a target for future retrievers.
    \item \textit{BM25 remains a robust baseline.} BM25 outperforms other baselines like DeepCT, DPR, SimCSE, and SPARTA while being significantly faster.
    \item \textit{Combine across families, not within them.} Not all pairs of retrievers are equally complementary: dense instruction-tuned models have higher hard-query overlap, so fusing two of them adds little, whereas cross-family pairs overlap far less. Candidates for fusion should therefore be selected on low measured overlap rather than on architectural difference, which our results show to be a poor proxy.
\end{itemize}

\textbf{Reproducibility.} In our experiments, we encountered substantial challenges affecting reproducibility. First, free-text prompt-based models tend to be highly sensitive to the wording of prompts, typically requiring dataset-specific prompts to obtain maximal performance. While this helps explain some of the discrepancies between our results and those in the literature especially for dense retrievers, it is problematic for practitioners for obvious reasons.
Second, inconsistencies between published papers and the authors' online resources are \textit{abundant}. In some cases, hyperparameters or implementation details described in Hugging Face model cards do not fully match those reported in the corresponding papers. Similarly, ambiguity surrounding model checkpoints poses a recurring challenge: updated checkpoints, multiple versions of the same model, or mismatches between checkpoint names and paper descriptions make it difficult to determine which version was used to obtain published results. This ambiguity is further compounded for models that offer both standard and SentenceTransformer variants, where the relationship between implementations is not always clearly specified.
Third, reproducibility is also affected by differences in evaluation pipelines. The use of API-based LLMs for query or document expansion introduces dependencies that are difficult to control or replicate over time. Moreover, evaluation details are sometimes omitted, such as the inclusion or exclusion of document titles during indexing or retrieval. For instance, including titles substantially improves performance on MS MARCO \citep{lassance2023msmarco}, while for BEIR’s ArguAna dataset it can degrade results, in some cases substantially.
Finally, the multitude of re-implementations of sparse retrievers and the use of custom packages with specialized wrappers can introduce inefficiencies,  particularly in parallelized settings, leading to variations in results even for traditional retrievers.

\textbf{Implications.}
Our findings indicate that a wide range of topically diverse benchmarks is needed to fully assess retriever performance, as the performance on BEIR, on which many approaches are exclusively evaluated, does not generalize to all domains. Furthermore, this hints at substantial potential for combining small retrievers with good overall performance (e.g., GTE) with reranking (e.g., jina-reranker \citep{wang2025jina-reranker-v3}), and context pruning (e.g., Provence \citep{chirkova2025provence}) to try to match or outperform the best large retriever, while keeping query response times low. 
Similarly, one could improve the performance of the best model, NV-Embed-v2, even further by using a multi-stage retrieval pipeline.
We note that this recommendation concerns a property our evaluation only partially measures. All our metrics are computed at small cutoffs, and so they reward a retriever for putting a relevant document near the very top of its ranking. In a pipeline with a reranker, what matters instead is how many relevant documents appear anywhere within the candidate set the reranker receives, which is typically the top 100 or top 1000. Our results therefore show which retrievers rank well on their own, but identifying the best candidate generators for a reranking pipeline would require evaluating at these larger cutoffs, which we leave to future work.
However, our findings also show that minor retrieval gains can come at the cost of disproportionately long query response times for some models. Since such models are unlikely to be useful outside of niche applications, standardized query runtime evaluation for new retrievers should become the norm, rather than the outlier it currently is, to enable readers to properly assess cost-performance trade-offs.

\textbf{Limitations.} 
Despite the breadth of our selection, a key limitation lies in the scope of models, datasets, and retrieval settings examined. Our experiments focus primarily on highly cited retrievers and exclude generative retrievers \citep{li2025survey}, hybrid retrievers \citep{bruch2024hybrid}, and reasoning retrievers \citep{shao2025reasonir}. The same applies to the wider expansion design space: the higher-cost families \citep{li2026queryexpansion}, such as interactive multi-round expansion, are characterized there as selective rather than default deployment strategies, and we leave their systematic evaluation to future work. We also excluded different variants of retrievers (e.g., E5$_\text{mistral-7b}$, BGE-M3, Qwen3 Embedding 8B) and less widely adopted models. While we evaluate a diverse set of benchmarks, our analysis does not cover specialized retrieval datasets, less established datasets, and multilingual datasets. These exclusions reflect necessary scope boundaries rather than conceptual gaps. The space of retrievers, model variants, and datasets is large and continuously expanding, and a comprehensive evaluation of all configurations is beyond the scope of any single paper. The selected models and datasets are therefore intended to be representative and maximize comparability to prior work, rather than exhaustive.

\section{Conclusion}

\textbf{Summary.} In this work, we evaluated 33 retrievers across seven datasets using a framework emphasizing fair comparisons. Newer instruction-tuned general-purpose models perform best overall, although their scores depend on the wording of instructions (e.g., due to dataset specificity) and on evaluation choices (e.g., the inclusion of titles), with effects varying across datasets and retrievers. Furthermore,  the efficiency of sparse retrievers increases with the number of documents and retrievers with LLM-backbones are on average $5\times$ slower than other retrievers. Additionally, our results show that retrievers largely struggle with distinct subsets of queries, indicating potential for hybrid models and multi-stage pipelines. Finally, we documented the (at times substantial) difficulties that we encountered in replicating reported results due to a lack of checkpoint and documentation availability, calling for greater care in how retrievers are shared, documented, evaluated, and compared.

\textbf{Future Work.} In our experiments, we found that the effectiveness of instruction-following retrievers is affected by prompt choice and dataset-specific prompt alignment, a finding that is consistent with prior work on the prompt sensitivity of large language models \citep{errica2025what, sclar2024quantifying}. Thus, an important direction for future work should include the systematic investigation of prompt sensitivity across retriever architectures, datasets, and task formulations, with an emphasis on understanding robustness, generalization, and the interaction between prompt design and retrieval performance.

\section*{Ethical Considerations}
As our work describes a benchmarking study of previously published approaches in the literature, we see no negative societal effects or potential for abuse of our work that would warrant ethical consideration.

\section*{Acknowledgements}
The authors acknowledge support by the state of Baden-Württemberg through bwHPC.

\bibliographystyle{ACM-Reference-Format-etal}
\bibliography{references}

\section*{Appendix}

\setcounter{section}{0}
\setcounter{table}{0}
\setcounter{figure}{0}

\renewcommand{\thesection}{A.\arabic{section}}
\renewcommand{\thetable}{A.\arabic{table}}
\renewcommand{\thefigure}{A.\arabic{figure}}

\input{appendix}

\end{document}

%% file: appendix.tex
This appendix provides supplementary material supporting the experiments presented in the main text. Appendix \ref{appendix-a} contains additional information about the evaluated models, datasets, and experimental setup. Appendix \ref{appendix-b} presents additional evaluation results that complement the analyses discussed in Section \ref{sec:results}.

\section{Additional Information}
\label{appendix-a}

The first section provides supplementary information about the models, datasets, and experimental setup discussed in the main text. Table \ref{tab-a:retrievers} summarizes the model descriptions of all retrievers in Section \ref{sec:retrievers}. Table \ref{tab-a:dataset-sizes} lists the exact sizes of the datasets used in the experiments introduced in Section \ref{subsec:data}. This information was obtained from the respective original publications of the datasets.

\section{Additional Results}
\label{appendix-b}

The second section contains additional evaluation results that complement the analyses presented in Section \ref{sec:results}.

Table \ref{tab-a:retrieval-quality-zscores} reports the exact bootstrap $z$-scores ($\Delta_{\sigma_t}$) used to determine the color coding of the results in Table \ref{tab:retrieval-quality}. Each value indicates how many standard deviations the bootstrapped mean of each reproduced result deviates from the corresponding published result. While most reproduced results remain within two standard deviations of the published values, several outliers can be observed. For example, particularly large deviations occur for KaLM-Embedding (v2.5) on BEIR ($\Delta_{\sigma_t}=-21.475$), SPARTA on BEIR ($\Delta_{\sigma_t}=-19.469$), and DeepCT on MS MARCO ($\Delta_{\sigma_t}=-16.538$), where the results from my experiments fall substantially below the published results.

Tables \ref{tab-a:msmarco-trec-results}--\ref{tab-a:instructir-followir-results} provide additional evaluation results. Specifically, Table \ref{tab-a:msmarco-trec-results} reports the performance of all retrievers on the in-domain datasets using Recall$@1000$. On MS MARCO, NV-Embed-v2 achieves the highest Recall$@1000$, indicating that it retrieves the relevant document within the top 1000 results more frequently than other models. However, SPLADE-v3 often ranks the relevant document higher, which explains its stronger performance on ranking-based metrics such as MRR. For TREC-DL 2019, the overall pattern remains similar to the main results table, with NV-Embed-v2 achieving the highest Recall$@1000$, tied with \textsc{GritLM}. In contrast, on TREC-DL 2020 the pattern reverses compared to MS MARCO, with SPLADE-v3 achieving a higher Recall$@1000$ than NV-Embed-v2.

Table \ref{tab-a:beir-results} presents disaggregated results for the individual datasets within the BEIR benchmark, where the values for CQADupStack are averages across the 12 CQADupStack datasets. While NV-Embed-v2 achieves the best performance on five of the fourteen datasets, GTE (large v1.5) performs best on six datasets. Despite not being the top model on the majority of datasets, NV-Embed-v2 generally achieves competitive scores when it is not the best-performing model. One notable exception is the Climate-FEVER dataset, where GTE (large v1.5) substantially outperforms NV-Embed-v2. This difference may be related to the difficulty of the dataset and the fact that dataset-specific instructions were not used, which may affect more challenging datasets to a greater extent.

Table \ref{tab-a:lotte-results} reports disaggregated results for the LoTTE benchmark, distinguishing between the Search and Forum subsets. The results largely mirror the patterns observed in the main results table, with NV-Embed-v2 achieving the best performance on both subsets.

Finally, Table \ref{tab-a:instructir-followir-results} provides disaggregated results for the datasets included in the \textsc{FollowIR} benchmark. Compared to the aggregated results in the main table, the performance patterns differ more strongly across individual datasets. In terms of the averaged score, the overall best-performing model \textsc{GritLM} achieves the highest scores on Robust04 (MAP) and Core17 (MAP), while NV-Embed-v2 performs best on News21 (nDCG$@5$). When considering $p$-MRR, the overall best-performing model SPARTA achieves the highest score only on Robust04, while SimCSE performs best on News21 and \textsc{GritLM} on Core17.

\begin{figure*}[h!]
    \centering
    \includegraphics[width=\linewidth]{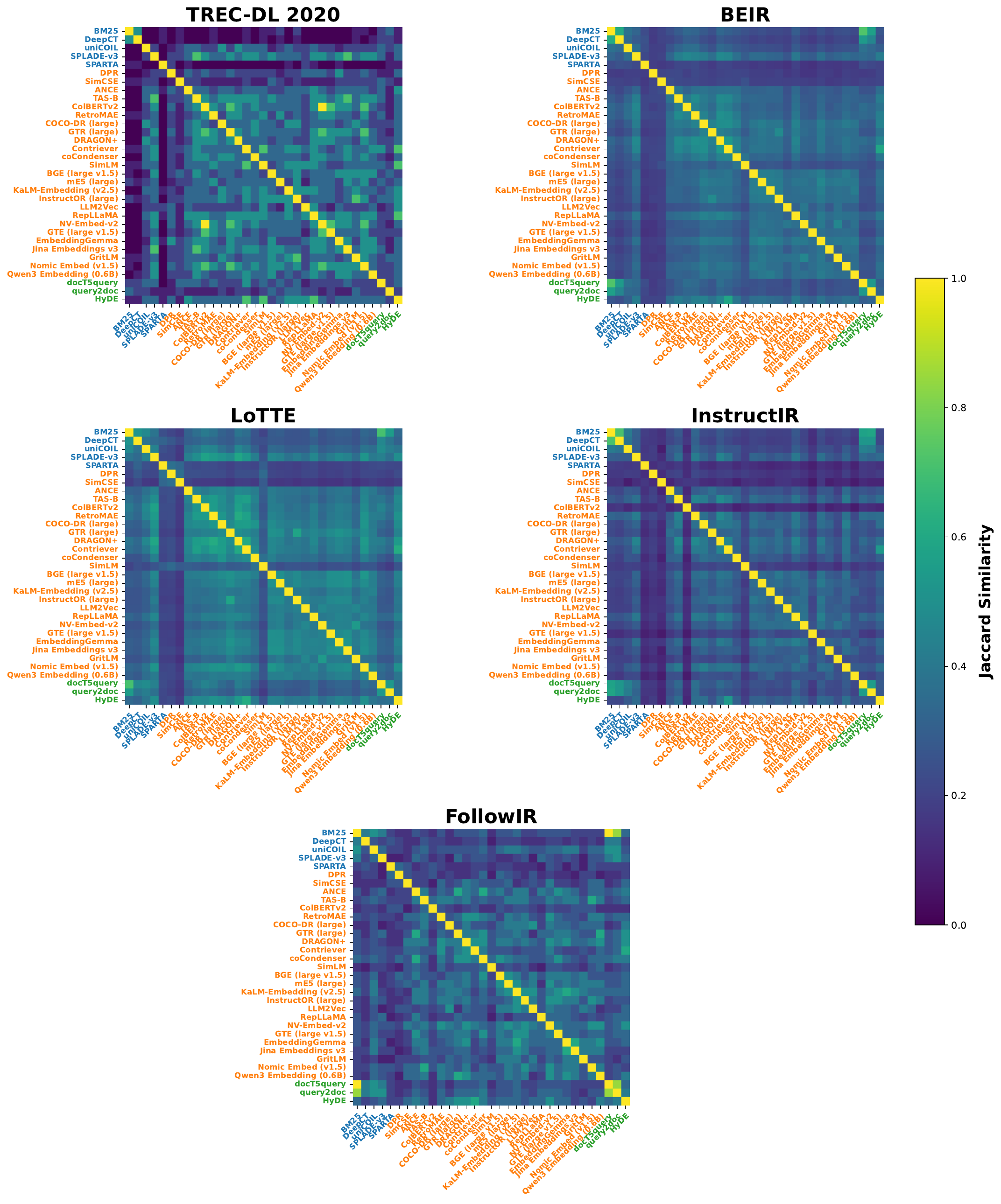}
    \caption{Jaccard similarities of hard-query sets. Pairwise Jaccard similarity between models based on the overlap of their hard-query sets for TREC-DL 2020 and BEIR (top), LoTTE and \textsc{InstructIR} (middle), and \textsc{FollowIR} (bottom), grouped by model family and using the hardest 10\% of queries per dataset.}
    \label{fig-a:hard-queries}
\end{figure*}

Figure \ref{fig-a:hard-queries} presents a plot with the worst-performing queries heatmaps for all retrievers on BEIR, LoTTE, \textsc{InstructIR}, and \textsc{FollowIR}. As discussed in Section \ref{sec:failure-point-ana}, the patterns in Figure \ref{fig-a:hard-queries} are similar to those in Figure \ref{fig:hard-queries}.

\begin{landscape}
\begin{table}[h]
    \centering
    \footnotesize
    \caption{An overview of retrievers by type. Note: BoW = Bag of Words; MSE = Mean Squared Error; MAE = Masked Auto-Encoder; MRL = Matryoshka Representation Learning; MLM = Masked Language Modeling; KL = Kullback-Leibler.}
    \vspace{5pt}
    \begin{tabular}{lllp{5cm}l}
        \toprule
        \textbf{Retriever} & \textbf{Type} & \textbf{Backbone} & \textbf{Training} & \textbf{Mechanism} \\
        \midrule
        BM25 & Sparse, BOW-based & -- & -- & Probabilistic relevance estimation using term frequency \\
        DeepCT & Learned Sparse & BERT & MSE loss & Contextualized term importance; BM25 retrieval \\
        SPARTA & Learned Sparse & BERT & Cross-entropy loss & Token-level representation; dot-product + max pooling; inverted index \\
        uniCOIL & Learned Sparse & BERT & Contrastive & Relevance scoring of exact lexical matches; max pooling \\
        SPLADE-v3 & Learned Sparse & coCondenser & Contrastive \& knowledge distillation loss (MarginMSE \& KL-divergence loss) & Expansion-aware term weights; log-transformation; max pooling \\
        \midrule
        DPR & Dense, Embedding-based & BERT (2) & Contrastive & Dual-encoder; dot product + \texttt{[CLS]} pooling \\
        ColBERTv2 & Dense, Embedding-based & BERT & Cross-entropy \& knowledge distillation loss & Late interaction; MaxSim pooling; indexing with $k$-means clustering \\
        TAS-B & Dense, Embedding-based & BERT & Contrastive \& knowledge distillation loss (Margin-MSE loss) & Dual-teacher supervision; topic-aware batch sampling \\
        ANCE & Dense, Embedding-based & BERT & Contrastive w/ hard negatives & Asynchronous index updating; dynamic corpus-level negatives \\
        SimCSE & Dense, Embedding-based & BERT, RoBERTa & Contrastive (unsupervised \& supervised) & Use of dropout during training \\
        RetroMAE & Dense, Embedding-based & BERT & MAE (reconstruction loss) & Asymmetric encoder-decoder \\
        coCondenser & Dense, Embedding-based & BERT & Cross-entropy, contrastive & Corpus-aware contrastive pretraining; improved \texttt{[CLS]} representations \\
        Contriever & Dense, Embedding-based & BERT & Contrastive w/ MoCo \& random cropping & Unsupervised training; momentum queue negatives \\
        COCO-DR (large) & Dense, Embedding-based & BERT & Contrastive (iDRO) & Continuous target-corpus pretraining + robust fine-tuning \\
        \textsc{Dragon}+ & Dense, Embedding-based & BERT, RoBERTa & Contrastive & Diverse query \& label augmentation; progressive teacher supervision \\
        mE5 (large) & General-purpose embedder & BERT, Mistral 7B & Contrastive (weakly supervised) & Multilingual embeddings; mean/last-token pooling \\
        GTE (large v1.5) & General-purpose embedder & BERT & Contrastive (improved bi-directional) & Large-scale unsupervised pretrain + supervised finetune \\
        BGE (large v1.5) & General-purpose embedder& BERT-style Transformer & MAE (reconstruction loss) \& contrastive & Multi-task fine-tuning; \texttt{[CLS]} pooling \\
        Jina Embeddings v3 & General-purpose embedder & XLM-RoBERTa & Contrastive (bi-directional) & Mean \& MaxSim pooling; multilingual embeddings; MRL \\
        Nomic Embed (v1.5) & General-purpose embedder & Nomic-BERT & MLM \& contrastive (weakly sup \& sup.) & 8k context; task-specific prefixes \\
        GTR (large) & Dense, Embedding-based & T5 & Contrastive (bi-directional) & Dual-encoder; cosine similarity \\
        \textsc{InstructOR} (large) & Instruction-tuned & GTR & Contrastive (bi-directional) & Instruction-tuned embeddings \\
        LLM2Vec & Instruction-tuned & LLaMA & MNTP + unsupervised SimCSE & Bidirectional attention; then contrastive sentence embedding \\
        RepLLaMA & Instruction-tuned & LLaMA & Contrastive & Last-token pooling; dot product \\
        GritLM & General-purpose embedder & Mistral 7B & Contrastive \& MLM & Instruction-tuning via GRIT and dual modes via instruction prefix \\
        Qwen3 Embedding (0.6B) & General-purpose embedder & Qwen3 & Multi-stage & Instruction-aware; last-token pooling; MRL dimensions \\
        NV-Embed-v2 & General-purpose embedder & Mistral 7B & Contrastive (two-stage instruction-tuning) & Bidirectional attention (no causal mask); latent attention pooling \\
        EmbeddingGemma & General-purpose embedder & Gemma 3 & Multi-stage & Complex training pipeline; task-specific prompts \\
        KaLM-Embedding (v2.5) & General-purpose embedder & Qwen2 0.5B & Multi-stage & Task-specific prompts; high-quality synthetic training data \\
        \midrule
        docT5query & Sparse, document expansion & T5 & -- & Augment documents with synthetic queries; BM25 retrieval \\
        query2doc & Sparse, query expansion & GPT-4, BM25 & -- & Augment queries with synthetic documents; any retrieval \\
        HyDE & Dense, query expansion & InstructGPT, Contriever & -- & Augment queries with synthetic documents; dense retrieval \\
        \bottomrule
    \end{tabular}
    \label{tab-a:retrievers}
\end{table}
\end{landscape}

\begin{table*}[h!]
    \centering
    \caption{Number of queries and documents for all evaluation datasets. \textsuperscript{*}Total across all 12 domains; \textsuperscript{\textdagger}Not publicly available.}
    \normalsize
    \begin{tabular}{lrr}
        \toprule
        \textbf{Dataset} & \textbf{\#Queries} & \textbf{\#Documents} \\
        \midrule
        \textbf{MS MARCO (passage, dev)} & 6{,}980 & 8{,}841{,}823 \\
        \midrule
        \multicolumn{3}{l}{\textbf{TREC DL (passage)}} \\
        \midrule
        TREC DL 2019 & 43 & 8{,}841{,}823 \\
        TREC DL 2020 & 54 & 8{,}841{,}823 \\
        \midrule
        \multicolumn{3}{l}{\textbf{BEIR}} \\
        \midrule
        TREC-COVID & 50 & 171{,}332 \\
        NFCorpus & 323 & 3{,}633 \\
        BioASQ\textsuperscript{\textdagger} & 500 & 14{,}914{,}602 \\
        NQ & 3{,}452 & 2{,}681{,}468 \\
        HotpotQA & 7{,}405 & 5{,}233{,}329 \\
        FiQA-2018 & 648 & 57{,}638 \\
        Signal-1M\textsuperscript{\textdagger} & 97 & 2{,}866{,}316 \\
        TREC-NEWS\textsuperscript{\textdagger} & 57 & 594{,}977 \\
        Robust04\textsuperscript{\textdagger} & 249 & 528{,}155 \\
        ArguAna & 1{,}406 & 8{,}674 \\
        Touché-2020 & 49 & 382{,}545 \\
        CQADupstack\textsuperscript{*} & 13{,}145 & 457{,}199 \\
        Quora & 10{,}000 & 522{,}931 \\
        DBPedia-entity & 400 & 4{,}635{,}922 \\
        SCIDOCS & 1{,}000 & 25{,}657 \\
        FEVER & 6{,}666 & 5{,}416{,}568 \\
        Climate-FEVER & 1{,}535 & 5{,}416{,}535 \\
        SciFact & 300 & 5{,}183 \\
        \midrule
        \multicolumn{3}{l}{\textbf{LoTTE (pooled)}} \\
        \midrule
        Search & 3{,}869 & 2{,}819{,}103 \\
        Forum & 10{,}025 & 2{,}819{,}103 \\
        \midrule
        \textbf{\textsc{InstructIR}} & 9{,}906 & 16{,}072 \\
        \midrule
        \multicolumn{3}{l}{\textbf{\textsc{FollowIR}}} \\
        \midrule
        TREC News’21 & 32 & 30{,}900 \\
        TREC Core’17 & 20 & 19{,}900 \\
        TREC Robust’04 & 52 & 47{,}500 \\
        \bottomrule
    \end{tabular}
    \label{tab-a:dataset-sizes}
\end{table*}

\begin{landscape}
\begin{table}[h!]
    \centering
    \small
    \caption{Evaluation results --- bootstrap $z$-scores. S$@5$ = Success$@5$ and R$@10$ = Robustness$@10$. Score is the average across MAP$@1000$ and nDCG$@5$. Positive values indicate higher reproduced scores than reported in the literature.}
    \begin{tabular}{lccccccccc}
        \toprule
        \textbf{Dataset} $\rightarrow$ & \textbf{MS MARCO} & \textbf{TREC-DL 19} & \textbf{TREC-DL 20} & \textbf{BEIR} & \textbf{LoTTE} & \multicolumn{2}{c}{\textbf{InstructIR}} & \multicolumn{2}{c}{\textbf{FollowIR}} \\
        \cmidrule(lr){2-2} \cmidrule(lr){3-3} \cmidrule(lr){4-4} \cmidrule(lr){5-5} \cmidrule(lr){6-6} \cmidrule(lr){7-8} \cmidrule(lr){9-10}
        \textbf{Model} $\downarrow$ & \textbf{MRR$@10$} & \textbf{nDCG$@10$} & \textbf{nDCG$@10$} & \textbf{nDCG$@10$} & \textbf{S$@5$} & \textbf{R$@10$} & \textbf{nDCG$@10$} & \textbf{Score} & \textbf{$p$-MRR} \\
        \midrule
        \multicolumn{10}{l}{\textbf{Sparse Retrievers}}\\
        \midrule
        \addlinespace[2pt]
        \multicolumn{9}{l}{\textit{Lexical}} \\
        \addlinespace[1pt]
        BM25 & 0.000 & 0.000 & 0.000 & 0.000 & -0.110 & 1.084 & 1.843 & 1.859 & -0.110 \\
        \addlinespace[2pt]
        \multicolumn{9}{l}{\textit{Learned Sparse}} \\
        \addlinespace[1pt]
        DeepCT & -16.538 & {} & {} & 14.710 & {} & {} & {} & {} & {} \\
        SPARTA & {} & {} & {} & -19.469 & {} & {} & {} & {} & {} \\
        uniCOIL & -1.711 & -1.094 & {} & 1.302 & {} & {} & {} & {} & {} \\
        SPLADE-v3 & 0.204 & 0.059 & -0.074 & -2.114 & -0.596 & {} & {} & {} & {} \\
        \midrule
        \multicolumn{10}{l}{\textbf{Dense Retrievers}}\\
        \midrule
        \addlinespace[2pt]
        \multicolumn{9}{l}{\textit{Encoder-Only}} \\
        \addlinespace[1pt]
        DPR & {} & {} & {} & -1.053 & {} & {} & {} & {} & {} \\
        ColBERTv2 & 0.000 & 0.000 & -0.076 & -2.320 & 0.000 & 0.077 & 0.338 & {} & {} \\
        TAS-B & -1.502 & -0.373 & 0.067 & -0.193 & {} & {} & {} & {} & {} \\
        ANCE & -12.416 & -1.246 & {} & -1.296 & -9.411 & {} & {} & {} & {} \\
        SimCSE & {} & {} & {} & -3.221 & {} & {} & {} & {} & {} \\
        RetroMAE & -2.138 & -0.254 & 0.383 & 0.025 & -0.973 & {} & {} & {} & {} \\
        coCondenser & -9.268 & 0.028 & {} & -3.433 & -1.965 & {} & {} & {} & {} \\
        Contriever & -0.878 & -0.214 & 0.427 & -0.439 & -0.912 & 0.000 & -0.200 & -0.103 & 0.983 \\
        COCO-DR (large) & -1.958 & -0.372 & 0.033 & -4.085 & 4.414 & {} & {} & {} & {} \\
        \textsc{SimLM} & -15.037 & -1.739 & -1.541 & {} & {} & {} & {} & {} & {} \\
        \textsc{Dragon}+ & -1.684 & -0.260 & -0.101 & -0.786 & -0.933 & {} & {} & {} & {} \\
        mE5 (large) & -9.100 & {} & {} & 3.321 & {} & {} & {} & {} & {} \\
        GTE (large v1.5) & -0.908 & {} & {} & -0.770 & {} & {} & {} & {} & {} \\
        BGE (large v1.5) & -1.950 & {} & {} & -5.537 & {} & {} & {} & -0.480 & 0.308 \\
        Jina Embeddings v3 & -1.035 & {} & {} & -3.243 & {} & {} & {} & {} & {} \\
        Nomic Embed (v1.5) & -0.750 & {} & {} & -0.440 & {} & {} & {} & {} & {} \\
        \addlinespace[2pt]
        \multicolumn{9}{l}{\textit{Encoder-Decoder}} \\
        \addlinespace[1pt]
        GTR (large) & -2.303 & {} & {} & 0.775 & {} & 0.073 & 0.168 & {} & {} \\
        \textsc{InstructOR} (large) & {} & {} & {} & -2.302 & {} & -0.839 & -3.772 & {} & {} \\
        \addlinespace[2pt]
        \multicolumn{9}{l}{\textit{Decoder-Only}} \\
        \addlinespace[1pt]
        LLM2Vec & {} & {} & {} & {} & {} & {} & {} & {} & {} \\
        RepLLaMA & -2.671 & -0.067 & 0.237 & -4.232 & {} & -3.378 & -5.083 & -0.763 & 0.507 \\
        \textsc{GritLM} & {} & {} & {} & -3.558 & {} & {} & {} & 1.277 & 2.900 \\
        Qwen3 Embedding (0.6B) & {} & {} & {} & {} & {} & {} & {} & {} & {} \\
        NV-Embed-v2 & {} & {} & {} & -7.034 & {} & {} & {} & {} & {} \\
        EmbeddingGemma & {} & {} & {} & {} & {} & {} & {} & {} & {} \\
        KaLM-Embedding (v2.5) & {} & {} & {} & -21.475 & {} & {} & {} & {} & {} \\
        \midrule
        \multicolumn{10}{l}{\textbf{Expansion-Based Retrievers}}\\
        \midrule
        docT5query & -3.144 & -1.168 & -1.343 & -0.239 & {} & {} & {} & {} & {} \\
        query2doc & -2.638 & -0.724 & -1.057 & {} & {} & {} & {} & {} & {} \\
        HyDE & {} & 3.637 & 4.167 & {} & {} & {} & {} & {} & {} \\
        \bottomrule
    \end{tabular}
    \label{tab-a:retrieval-quality-zscores}
\end{table}
\end{landscape}

\begin{table*}[h]
\centering
\normalsize
\caption{Evaluation results --- extended results for MS MARCO and TREC DL (2019 \& 2020). R$@1000$ = Recall$@1000$. Results for MS MARCO (MRR$@10$), TREC-DL 2019 (nDCG$@10$), and TREC-DL 2020 (nDCG$@10$) were copied from Table \ref{tab:retrieval-quality} for comparisons. The best result per dataset is highlighted bold.}
\begin{tabular}{lcccccc}
\toprule
\textbf{Dataset $\rightarrow$} & \multicolumn{2}{c}{\textbf{MS MARCO}} & \multicolumn{2}{c}{\textbf{TREC DL 2019}} & \multicolumn{2}{c}{\textbf{TREC DL 2020}} \\
\cmidrule(lr){2-3} \cmidrule(lr){4-5} \cmidrule(lr){6-7}
\textbf{Model $\downarrow$} & {\textbf{MRR}@10} & {\textbf{R}@1000} & {\textbf{nDCG}@10} & {\textbf{R}@1000} & {\textbf{nDCG}@10} & {\textbf{R}@1000} \\
\midrule
\multicolumn{7}{l}{\textbf{Sparse Retrievers}}\\
\midrule
\addlinespace[2pt]
\multicolumn{7}{l}{\textit{Lexical}} \\
\addlinespace[1pt]
BM25 & 0.185 & 0.857 & 0.512 & 0.741 & 0.477 & 0.722 \\
\addlinespace[2pt]
\multicolumn{7}{l}{\textit{Learned Sparse}} \\
\addlinespace[1pt]
DeepCT & 0.180 & 0.840 & 0.507 & 0.708 & 0.447 & 0.729 \\
SPARTA & 0.142 & 0.766 & 0.351 & 0.479 & 0.285 & 0.506 \\
uniCOIL & 0.307 & 0.920 & 0.648 & 0.740 & 0.642 & 0.723 \\
SPLADE-v3 & \textbf{0.403} & 0.987 & 0.725 & 0.830 & 0.752 & \textbf{0.844} \\
\midrule
\multicolumn{7}{l}{\textbf{Dense Retrievers}}\\
\midrule
\addlinespace[2pt]
\multicolumn{7}{l}{\textit{Encoder-Only}} \\
\addlinespace[1pt]
DPR & 0.106 & 0.707 & 0.297 & 0.453 & 0.270 & 0.461 \\
ColBERTv2 & 0.397 & 0.983 & 0.746 & 0.840 & 0.750 & 0.842 \\
TAS-B & 0.340 & 0.976 & 0.704 & 0.783 & 0.688 & 0.807 \\
ANCE & 0.275 & 0.934 & 0.592 & 0.647 & 0.590 & 0.662 \\
SimCSE & 0.067 & 0.562 & 0.251 & 0.364 & 0.190 & 0.383 \\
RetroMAE & 0.344 & 0.974 & 0.679 & 0.787 & 0.724 & 0.783 \\
coCondenser & 0.340 & 0.972 & 0.685 & 0.748 & 0.674 & 0.764 \\
Contriever & 0.337 & 0.980 & 0.670 & 0.792 & 0.675 & 0.803 \\
COCO-DR (large) & 0.349 & 0.982 & 0.729 & 0.798 & 0.698 & 0.786 \\
\textsc{SimLM} & 0.342 & 0.945 & 0.650 & 0.721 & 0.640 & 0.730 \\
\textsc{Dragon}+ & 0.382 & 0.987 & 0.736 & 0.798 & 0.720 & 0.820 \\
mE5 (large) & 0.300 & 0.975 & 0.633 & 0.797 & 0.673 & 0.787 \\
GTE (large v1.5) & 0.361 & 0.982 & 0.708 & 0.807 & 0.729 & 0.789 \\
BGE (large v1.5) & 0.353 & 0.976 & 0.697 & 0.790 & 0.698 & 0.754 \\
Jina Embeddings v3 & 0.341 & 0.976 & 0.699 & 0.737 & 0.683 & 0.743 \\
Nomic Embed (v1.5) & 0.360 & 0.977 & 0.716 & 0.777 & 0.701 & 0.758 \\
\addlinespace[2pt]
\multicolumn{7}{l}{\textit{Encoder-Decoder}} \\
\addlinespace[1pt]
GTR (large) & 0.368 & 0.988 & 0.713 & 0.768 & 0.687 & 0.755 \\
\textsc{InstructOR} (large) & 0.343 & 0.982 & 0.645 & 0.752 & 0.668 & 0.745 \\
\addlinespace[2pt]
\multicolumn{7}{l}{\textit{Decoder-Only}} \\
\addlinespace[1pt]
LLM2Vec & 0.317 & 0.978 & 0.697 & 0.819 & 0.685 & 0.793 \\
RepLLaMA & 0.399 & 0.992 & 0.741 & 0.832 & 0.728 & 0.794 \\
\textsc{GritLM} & 0.356 & 0.989 & 0.721 & \textbf{0.874} & 0.730 & 0.838 \\
Qwen3 Embedding (0.6B) & 0.309 & 0.977 & 0.665 & 0.810 & 0.661 & 0.769 \\
NV-Embed-v2 & 0.389 & \textbf{0.993} & \textbf{0.753} & \textbf{0.874} & \textbf{0.765} & 0.831 \\
EmbeddingGemma & 0.317 & 0.979 & 0.718 & 0.828 & 0.714 & 0.809 \\
KaLM-Embedding (v2.5) & 0.265 & 0.968 & 0.568 & 0.761 & 0.582 & 0.757 \\
\midrule
\multicolumn{7}{l}{\textbf{Expansion-Based Retrievers}}\\
\midrule
docT5query & 0.246 & 0.919 & 0.602 & 0.787 & 0.576 & 0.780 \\
query2doc & 0.204 & 0.917 & 0.634 & 0.824 & 0.598 & 0.825 \\
HyDE & 0.302 & 0.977 & 0.728 & 0.847 & 0.716 & 0.843 \\
\bottomrule
\end{tabular}
\label{tab-a:msmarco-trec-results}
\end{table*}

\clearpage
\begin{landscape}
\begin{table}[h]
\centering
\small
\caption{Evaluation results --- extended results for BEIR (nDCG$@10$). The best result per dataset is highlighted bold.}
\begin{tabular}{lcccccccccccccc}
\toprule
\textbf{Dataset $\rightarrow$} & \multirow{2}{*}{\textbf{trec-covid}} & \multirow{2}{*}{\textbf{nfcorpus}} & \multirow{2}{*}{\textbf{nq}} & \multirow{2}{*}{\textbf{hotpotqa}} & \multirow{2}{*}{\textbf{fiqa}} & \multirow{2}{*}{\textbf{arg}} & \multirow{2}{*}{\textbf{touche}} & \multirow{2}{*}{\textbf{cqa}} & \multirow{2}{*}{\textbf{quora}} & \multirow{2}{*}{\textbf{dbpedia}} & \multirow{2}{*}{\textbf{scidocs}} & \multirow{2}{*}{\textbf{fever}} & \multirow{2}{*}{\textbf{climate-fever}} & \multirow{2}{*}{\textbf{scifact}} \\
\textbf{Model $\downarrow$} &  &  &  &  &  &  &  &  &  &  &  &  &  &  \\
\midrule
\multicolumn{15}{l}{\textbf{Sparse Retrievers}}\\
\midrule
\addlinespace[2pt]
\multicolumn{15}{l}{\textit{Lexical}} \\
\addlinespace[1pt]
BM25 & 0.595 & 0.322 & 0.305 & 0.633 & 0.236 & 0.397 & \textbf{0.442} & 0.302 & 0.789 & 0.318 & 0.149 & 0.651 & 0.165 & 0.679 \\
\addlinespace[2pt]
\multicolumn{15}{l}{\textit{Learned Sparse}} \\
\addlinespace[1pt]
DeepCT & 0.426 & 0.311 & 0.224 & 0.548 & 0.188 & 0.458 & 0.298 & 0.275 & 0.761 & 0.286 & 0.125 & 0.570 & 0.103 & 0.680 \\
SPARTA & 0.464 & 0.255 & 0.260 & 0.378 & 0.134 & 0.348 & 0.133 & 0.205 & 0.358 & 0.190 & 0.112 & 0.428 & 0.057 & 0.518 \\
uniCOIL & 0.661 & 0.335 & 0.422 & 0.666 & 0.280 & 0.381 & 0.319 & 0.302 & 0.685 & 0.348 & 0.144 & 0.816 & 0.192 & 0.687 \\
SPLADE-v3 & 0.733 & 0.363 & 0.586 & 0.693 & 0.381 & 0.506 & 0.310 & 0.345 & 0.814 & 0.451 & 0.157 & 0.806 & 0.250 & 0.716 \\
\midrule
\multicolumn{15}{l}{\textbf{Dense Retrievers}}\\
\midrule
\addlinespace[2pt]
\multicolumn{15}{l}{\textit{Encoder-Only}} \\
\addlinespace[1pt]
DPR & 0.291 & 0.187 & 0.465 & 0.384 & 0.097 & 0.154 & 0.111 & 0.144 & 0.349 & 0.258 & 0.073 & 0.549 & 0.146 & 0.308 \\
ColBERTv2 & 0.727 & 0.334 & 0.563 & 0.675 & 0.356 & 0.460 & 0.260 & 0.367 & 0.853 & 0.449 & 0.154 & 0.786 & 0.179 & 0.689 \\
TAS-B & 0.481 & 0.319 & 0.463 & 0.584 & 0.297 & 0.427 & 0.162 & 0.314 & 0.834 & 0.385 & 0.148 & 0.700 & 0.228 & 0.643 \\
ANCE & 0.523 & 0.220 & 0.386 & 0.480 & 0.275 & 0.440 & 0.208 & 0.318 & 0.841 & 0.279 & 0.124 & 0.675 & 0.293 & 0.518 \\
SimCSE & 0.364 & 0.123 & 0.111 & 0.215 & 0.137 & 0.454 & 0.115 & 0.190 & 0.799 & 0.129 & 0.070 & 0.182 & 0.163 & 0.369 \\
RetroMAE & 0.772 & 0.308 & 0.518 & 0.635 & 0.314 & 0.433 & 0.237 & 0.317 & 0.842 & 0.390 & 0.150 & 0.773 & 0.232 & 0.653 \\
coCondenser & 0.674 & 0.315 & 0.471 & 0.544 & 0.250 & 0.337 & 0.160 & 0.313 & 0.856 & 0.357 & 0.136 & 0.441 & 0.130 & 0.601 \\
Contriever & 0.596 & 0.328 & 0.498 & 0.638 & 0.330 & 0.446 & 0.204 & 0.345 & 0.865 & 0.413 & 0.165 & 0.758 & 0.237 & 0.677 \\
COCO-DR (large) & 0.808 & 0.355 & 0.543 & 0.633 & 0.319 & 0.455 & 0.216 & 0.374 & 0.868 & 0.408 & 0.173 & 0.749 & 0.231 & 0.719 \\
\textsc{SimLM} & 0.408 & 0.224 & 0.385 & 0.551 & 0.179 & 0.342 & 0.097 & 0.295 & 0.857 & 0.331 & 0.119 & 0.317 & 0.087 & 0.494 \\
\textsc{Dragon}+ & 0.742 & 0.336 & 0.536 & 0.660 & 0.354 & 0.482 & 0.247 & 0.355 & 0.871 & 0.415 & 0.159 & 0.772 & 0.224 & 0.683 \\
mE5 (large) & 0.827 & 0.364 & 0.591 & 0.688 & 0.435 & 0.547 & 0.315 & 0.424 & 0.873 & 0.418 & 0.184 & 0.822 & 0.257 & 0.711 \\
GTE (large v1.5) & 0.772 & 0.370 & 0.559 & 0.682 & 0.608 & \textbf{0.720} & 0.216 & 0.423 & \textbf{0.897} & 0.463 & \textbf{0.263} & \textbf{0.939} & \textbf{0.485} & \textbf{0.827} \\
BGE (large v1.5) & 0.644 & 0.368 & 0.518 & 0.743 & 0.443 & 0.636 & 0.242 & 0.417 & 0.884 & 0.422 & 0.209 & 0.846 & 0.284 & 0.735 \\
Jina Embeddings v3 & 0.777 & 0.366 & 0.643 & 0.646 & 0.466 & 0.578 & 0.263 & 0.360 & 0.748 & 0.411 & 0.199 & 0.891 & 0.424 & 0.725 \\
Nomic Embed (v1.5) & 0.821 & 0.346 & 0.597 & 0.727 & 0.373 & 0.482 & 0.300 & 0.397 & 0.859 & 0.440 & 0.187 & 0.863 & 0.413 & 0.703 \\
\addlinespace[2pt]
\multicolumn{15}{l}{\textit{Encoder-Decoder}} \\
\addlinespace[1pt]
GTR (large) & 0.567 & 0.326 & 0.551 & 0.578 & 0.424 & 0.521 & 0.283 & 0.366 & 0.868 & 0.396 & 0.155 & 0.727 & 0.269 & 0.634 \\
\textsc{InstructOR} (large) & 0.464 & 0.339 & 0.485 & 0.569 & 0.453 & 0.556 & 0.206 & 0.432 & 0.875 & 0.370 & 0.185 & 0.748 & 0.277 & 0.639 \\
\addlinespace[2pt]
\multicolumn{15}{l}{\textit{Decoder-Only}} \\
\addlinespace[1pt]
LLM2Vec & 0.771 & 0.384 & 0.537 & 0.645 & 0.503 & 0.562 & 0.221 & 0.402 & 0.875 & 0.434 & 0.194 & 0.780 & 0.251 & 0.759 \\
RepLLaMA & \textbf{0.846} & 0.379 & 0.624 & 0.686 & 0.448 & 0.482 & 0.269 & 0.371 & 0.865 & 0.438 & 0.180 & 0.833 & 0.314 & 0.742 \\
\textsc{GritLM} & 0.756 & 0.409 & \textbf{0.701} & 0.789 & 0.602 & 0.593 & 0.270 & 0.476 & 0.871 & 0.452 & 0.229 & 0.832 & 0.228 & 0.790 \\
Qwen3 Embedding (0.6B) & 0.742 & 0.361 & 0.514 & 0.614 & 0.430 & 0.675 & 0.238 & 0.443 & 0.821 & 0.381 & 0.229 & 0.878 & 0.382 & 0.697 \\
NV-Embed-v2 & 0.839 & \textbf{0.445} & 0.640 & \textbf{0.833} & \textbf{0.649} & 0.696 & 0.362 & \textbf{0.489} & 0.886 & \textbf{0.494} & 0.226 & 0.894 & 0.317 & 0.803 \\
EmbeddingGemma & 0.805 & 0.389 & 0.649 & 0.707 & 0.469 & 0.670 & 0.369 & 0.400 & 0.864 & 0.441 & 0.188 & 0.829 & 0.277 & 0.783 \\
KaLM-Embedding (v2.5) & 0.586 & 0.336 & 0.326 & 0.673 & 0.421 & 0.604 & 0.121 & 0.431 & 0.886 & 0.235 & 0.133 & 0.798 & 0.309 & 0.727 \\
\midrule
\multicolumn{15}{l}{\textbf{Expansion-Based Retrievers}}\\
\midrule
docT5query & 0.646 & 0.326 & 0.353 & 0.657 & 0.262 & 0.416 & 0.409 & 0.319 & 0.782 & 0.335 & 0.153 & 0.695 & 0.171 & 0.678 \\
query2doc & 0.683 & 0.359 & 0.425 & 0.666 & 0.241 & 0.398 & 0.422 & 0.269 & 0.760 & 0.376 & 0.154 & 0.766 & 0.237 & 0.713 \\
HyDE & 0.762 & 0.355 & 0.558 & 0.655 & 0.333 & 0.447 & 0.320 & 0.304 & 0.843 & 0.429 & 0.164 & 0.831 & 0.271 & 0.700 \\
\bottomrule
\end{tabular}
\label{tab-a:beir-results}
\end{table}
\end{landscape}

\begin{table*}[h]
\centering
\normalsize
\caption{Evaluation results --- extended results for LoTTE (Success$@5$). The best result per dataset is highlighted bold.}
\begin{tabular}{lcc}
\toprule
\textbf{Dataset $\rightarrow$} & \multicolumn{2}{c}{\textbf{Pooled}} \\
\cmidrule(lr){2-3}
\textbf{Model $\downarrow$} & {\textbf{Search}} & {\textbf{Forum}} \\
\midrule
\multicolumn{3}{l}{\textbf{Sparse Retrievers}}\\
\midrule
\addlinespace[2pt]
\multicolumn{3}{l}{\textit{Lexical}} \\
\addlinespace[1pt]
BM25 & 0.483 & 0.472 \\
\addlinespace[2pt]
\multicolumn{3}{l}{\textit{Learned Sparse}} \\
\addlinespace[1pt]
DeepCT & 0.387 & 0.390 \\
SPARTA & 0.367 & 0.321 \\
uniCOIL & 0.614 & 0.540 \\
SPLADE-v3 & 0.754 & 0.647 \\
\midrule
\multicolumn{3}{l}{\textbf{Dense Retrievers}}\\
\midrule
\addlinespace[2pt]
\multicolumn{3}{l}{\textit{Encoder-Only}} \\
\addlinespace[1pt]
DPR & 0.306 & 0.252 \\
ColBERTv2 & 0.720 & 0.631 \\
TAS-B & 0.627 & 0.516 \\
ANCE & 0.604 & 0.530 \\
SimCSE & 0.279 & 0.261 \\
RetroMAE & 0.663 & 0.582 \\
coCondenser & 0.614 & 0.514 \\
Contriever & 0.658 & 0.585 \\
COCO-DR (large) & 0.682 & 0.601 \\
\textsc{SimLM} & 0.516 & 0.409 \\
\textsc{Dragon}+ & 0.728 & 0.621 \\
mE5 (large) & 0.760 & 0.743 \\
GTE (large v1.5) & 0.808 & 0.780 \\
BGE (large v1.5) & 0.777 & 0.708 \\
Jina Embeddings v3 & 0.798 & 0.770 \\
Nomic Embed (v1.5) & 0.760 & 0.672 \\
\addlinespace[2pt]
\multicolumn{3}{l}{\textit{Encoder-Decoder}} \\
\addlinespace[1pt]
GTR (large) & 0.795 & 0.727 \\
\textsc{InstructOR} (large) & 0.783 & 0.754 \\
\addlinespace[2pt]
\multicolumn{3}{l}{\textit{Decoder-Only}} \\
\addlinespace[1pt]
LLM2Vec & 0.806 & 0.787 \\
RepLLaMA & 0.797 & 0.726 \\
\textsc{GritLM} & 0.844 & 0.862 \\
Qwen3 Embedding (0.6B) & 0.706 & 0.672 \\
NV-Embed-v2 & \textbf{0.874} & \textbf{0.878} \\
EmbeddingGemma & 0.808 & 0.741 \\
KaLM-Embedding (v2.5) & 0.671 & 0.619 \\
\midrule
\multicolumn{3}{l}{\textbf{Expansion-Based Retrievers}}\\
\midrule
docT5query & 0.566 & 0.519 \\
query2doc & 0.495 & 0.496 \\
HyDE & 0.700 & 0.616 \\
\bottomrule
\end{tabular}
\label{tab-a:lotte-results}
\end{table*}

\begin{table*}[h]
\centering
\normalsize
\caption{Evaluation results --- extended results for \textsc{InstructIR} and \textsc{FollowIR}. Results for \textsc{InstructIR}, both Robustness$@10$ and nDCG$@10$, were copied from Table \ref{tab:retrieval-quality} for comparisons. The best result per dataset is highlighted bold.}
\begin{tabular}{lcccccccc}
\toprule
\textbf{Dataset $\rightarrow$} & \multicolumn{2}{c}{\textbf{\textsc{InstructIR}}} & \multicolumn{6}{c}{\textbf{\textsc{FollowIR}}} \\
\cmidrule(lr){2-3} \cmidrule(lr){4-9}
& \multicolumn{2}{c}{\textbf{MS MARCO}} & \multicolumn{2}{c}{\textbf{Robust04}} & \multicolumn{2}{c}{\textbf{News21}} & \multicolumn{2}{c}{\textbf{Core17}} \\
\cmidrule(lr){2-3} \cmidrule(lr){4-5} \cmidrule(lr){6-7} \cmidrule(lr){8-9}
\textbf{Model $\downarrow$} & \textbf{Robustness}@10 & \textbf{nDCG}@10 & \textbf{MAP} & $p$-\textbf{MRR} & \textbf{nDCG}@5 & $p$-\textbf{MRR} & \textbf{MAP} & $p$-\textbf{MRR} \\
\midrule
\multicolumn{9}{l}{\textbf{Sparse Retrievers}}\\
\midrule
\addlinespace[2pt]
\multicolumn{9}{l}{\textit{Lexical}} \\
\addlinespace[1pt]
BM25 & 0.277 & 0.766 & 17.898 & 0.265 & 17.471 & -4.198 & 14.669 & -3.040 \\
\addlinespace[2pt]
\multicolumn{9}{l}{\textit{Learned Sparse}} \\
\addlinespace[1pt]
DeepCT & 0.282 & 0.772 & 13.769 & 0.145 & 14.937 & -2.887 & 10.974 & -4.579 \\
SPARTA & 0.076 & 0.542 & 5.319 & \textbf{9.243} & 11.735 & 2.814 & 6.160 & 6.715 \\
uniCOIL & 0.284 & 0.731 & 18.321 & -4.072 & 22.983 & -1.116 & 17.492 & 5.733 \\
SPLADE-v3 & 0.498 & 0.873 & 21.275 & -8.057 & 19.803 & -1.716 & 20.598 & 1.390 \\
\midrule
\multicolumn{9}{l}{\textbf{Dense Retrievers}}\\
\midrule
\addlinespace[2pt]
\multicolumn{9}{l}{\textit{Encoder-Only}} \\
\addlinespace[1pt]
DPR & 0.211 & 0.637 & 7.993 & 0.849 & 15.428 & -2.726 & 11.469 & 5.379 \\
ColBERTv2 & 0.142 & 0.686 & 16.440 & -2.012 & 22.681 & 0.000 & 13.403 & 2.165 \\
TAS-B & 0.428 & 0.825 & 17.238 & -6.852 & 20.465 & -3.646 & 17.639 & 0.868 \\
ANCE & 0.325 & 0.718 & 12.797 & -8.581 & 18.099 & -1.268 & 12.790 & 1.598 \\
SimCSE & 0.163 & 0.651 & 9.259 & -6.098 & 17.350 & \textbf{3.057} & 10.077 & 5.623 \\
RetroMAE & 0.435 & 0.835 & 16.736 & -8.711 & 18.413 & -2.921 & 18.631 & 2.469 \\
coCondenser & 0.367 & 0.744 & 11.492 & -6.644 & 19.080 & -2.551 & 10.607 & 5.501 \\
Contriever & 0.474 & 0.848 & 17.870 & -5.182 & 20.755 & -1.444 & 18.598 & -0.022 \\
COCO-DR (large) & 0.428 & 0.809 & 17.127 & -1.471 & 19.139 & -2.261 & 15.913 & 0.408 \\
\textsc{SimLM} & 0.271 & 0.712 & 6.466 & -5.639 & 4.280 & -5.934 & 4.749 & -1.755 \\
\textsc{Dragon}+ & 0.498 & 0.869 & 19.642 & -4.197 & 24.484 & -1.002 & 17.346 & -0.004 \\
mE5 (large) & 0.570 & 0.893 & 21.404 & -5.286 & 23.211 & 1.418 & 21.821 & 4.332 \\
GTE (large v1.5) & 0.582 & 0.892 & 22.416 & -2.879 & 22.177 & -0.696 & 20.579 & 3.321 \\
BGE (large v1.5) & 0.521 & 0.869 & 15.808 & -6.477 & 20.959 & 1.334 & 15.339 & -0.370 \\
Jina Embeddings v3 & 0.488 & 0.835 & 24.121 & -2.800 & 26.110 & 1.952 & 21.599 & 2.405 \\
Nomic Embed (v1.5) & 0.429 & 0.801 & 17.788 & -7.252 & 22.798 & 0.391 & 18.317 & 1.225 \\
\addlinespace[2pt]
\multicolumn{9}{l}{\textit{Encoder-Decoder}} \\
\addlinespace[1pt]
GTR (large) & 0.376 & 0.760 & 15.835 & -13.518 & 21.177 & -2.156 & 16.591 & -0.285 \\
\textsc{InstructOR} (large) & 0.562 & 0.887 & 17.602 & -6.057 & 25.638 & -2.627 & 17.757 & -0.729 \\
\addlinespace[2pt]
\multicolumn{9}{l}{\textit{Decoder-Only}} \\
\addlinespace[1pt]
LLM2Vec & 0.639 & 0.916 & 24.396 & 5.953 & 25.975 & 2.319 & 25.338 & 8.348 \\
RepLLaMA & 0.504 & 0.864 & 23.120 & -8.234 & 20.666 & -1.695 & 20.361 & 3.032 \\
\textsc{GritLM} & \textbf{0.753} & \textbf{0.951} & \textbf{29.053} & 5.032 & 25.855 & 1.543 & \textbf{27.785} & \textbf{10.036} \\
Qwen3 Embedding (0.6B) & 0.460 & 0.817 & 25.661 & 5.291 & 25.969 & 1.533 & 21.161 & 9.264 \\
NV-Embed-v2 & 0.677 & 0.928 & 26.878 & -3.525 & \textbf{27.150} & -0.322 & 26.372 & 2.781 \\
EmbeddingGemma & 0.602 & 0.906 & 22.216 & -2.538 & 25.401 & -0.646 & 23.880 & 6.314 \\
KaLM-Embedding (v2.5) & 0.553 & 0.873 & 23.723 & -4.463 & 23.827 & -1.486 & 20.525 & 1.426 \\
\midrule
\multicolumn{9}{l}{\textbf{Expansion-Based Retrievers}}\\
\midrule
docT5query & 0.349 & 0.799 & 18.671 & 0.417 & 16.513 & -3.986 & 15.276 & -3.166 \\
query2doc & 0.356 & 0.803 & 21.281 & -2.471 & 19.484 & -3.766 & 18.371 & -1.604 \\
HyDE & 0.464 & 0.836 & 19.574 & -0.743 & 22.417 & -1.690 & 21.974 & 0.200 \\
\bottomrule
\end{tabular}
\label{tab-a:instructir-followir-results}
\end{table*}

